\documentclass[aps,prb,a4paper,nobibnotes,reprint,superscriptaddress]{revtex4-1}
\usepackage{amssymb,amsmath}
\usepackage{graphicx}
\usepackage{color}
\usepackage[svgnames]{xcolor}
\usepackage{array}
\usepackage{natbib}
\usepackage{soul}
\usepackage{multirow}
\usepackage[normalem]{ulem}

\begin{document}

\title{Correlated Electronic Properties of Some Graphene Nanoribbons: A DMRG Study}

\date{\today}

\author{V. M. L. Durga Prasad Goli}
\email[Electronic mail:]{durgaprasad.vml@gmail.com}
\thanks{VMLDPG and SP have contributed equally to this work.}
\affiliation{Solid State and Structural Chemistry Unit, Indian Institute of Science, Bangalore-560012, India.}
\author{Suryoday Prodhan}
\email[Electronic mail: ]{suryodayp@sscu.iisc.ernet.in}
\affiliation{Solid State and Structural Chemistry Unit, Indian Institute of Science, Bangalore-560012, India.}
\author{Sumit Mazumdar}
\email[Electronic mail: ]{sumit@physics.arizona.edu}
\affiliation{Department of Physics, University of Arizona, Tucson, Arizona 85721, USA.}
\affiliation{College of Optical Sciences, University of Arizona, Tucson, Arizona 85721, USA.}
\author{S. Ramasesha}
\email[Electronic mail: ]{ramasesh@sscu.iisc.ernet.in}
\affiliation{Solid State and Structural Chemistry Unit, Indian Institute of Science, Bangalore-560012, India.}

\begin{abstract}

The significant electron-electron interactions that characterize the $\pi$-electrons of
graphene nanoribbons (GNRs) necessitate going beyond one-electron tight-binding description.
Existing theories of electron-electron interactions in
GNRs take into account one electron-one hole interactions accurately but miss higher order effects.
We report highly accurate density matrix renormalization group (DMRG) calculations 
of the ground state electronic structure, the relative energies
of the lowest one-photon versus two-photon excitations and the charge gaps in three narrow
graphene nanoribbons (GNRs) within the correlated Pariser-Parr-Pople model for $\pi$-conjugated systems.
We have employed the symmetrized DMRG method to investigate the zigzag nanoribbon 3-ZGNR and two armchair
nanoribbons 6-AGNR and 5-AGNR, respectively. We predict bulk magnetization of the ground state of 3-ZGNR, and
a large spin gap in 6-AGNR in their respective thermodynamic limits. Nonzero charge gaps and semiconducting behavior, 
with moderate to large exciting binding energies are found for all three nanoribbons,
in contradiction to the prediction of tight-binding theory. 
The lowest two-photon gap in 3-ZGNR vanishes in the
thermodynamic limit, while this gap is smaller than the one-photon gap in 5-AGNR. However, in 6-AGNR the 
one-photon gap is smaller than the two-photon gap and it is predicted to be fluorescent.

\end{abstract}

\pacs{71.27.+a,73.22.-f,78.67.-n}

\keywords{Strongly correlated electronic systems; Symmetrized DMRG calculation; Graphene nanoribbons; 
Low-lying electronic states.}

\maketitle

\section{Introduction}
\label{intro}

Carbon has come to the fore in the last few decades for many exciting electronic and magnetic 
properties with the discovery of buckyball in the eighties to carbon nanotubes in the nineties
to graphene and graphene nanoribbons in the last decade \cite{saito, dresselhaus, geim}. In all these
systems, we can assume that the carbon atom is in sp$^2$ hybridization and hence,
these systems belong to the class of $\pi$-conjugated carbon systems. In recent years, graphene nanoribbons 
(GNRs) have attracted considerable attention because of their exotic electronic properties and plausible 
applications in nanoelectronics \cite{neto, allen, sarma, Schwierz, morris}. 
GNRs with different widths can be made using various techniques like mechanical cutting 
of exfoliated graphenes or by patterning epitaxially grown graphenes \cite{berger, chen, han, tapaszto, datta1, li}.
GNRs are quasi one-dimensional forms of graphene, which exhibit exciting electronic properties because of the 
confinement of electrons in low dimension \cite{neto, fujita, nakada, ezawa, Wakabayashi}. These electronic 
properties also depend  crucially on the geometry of the edges of the ribbons. The GNRs are classified into 
two types based on the edge structures, namely, zigzag and armchair GNRs (ZGNRs and AGNRs respectively).
Within the one-band tight-binding theory, ZGNRs are predicted to be metallic with zero bandgap, while AGNRs can be
either semiconducting or metallic, depending upon their width \cite{fujita, nakada, ezawa, brey};
AGNRs with 3p+2 (with integer p) dimer bonds between nearest neighbor carbon atoms
across the ribbon width are metallic and others are semiconducting \cite{ezawa}. 

\begin{figure}[tbp]
\begin{center} 
\includegraphics[width=8.5cm]{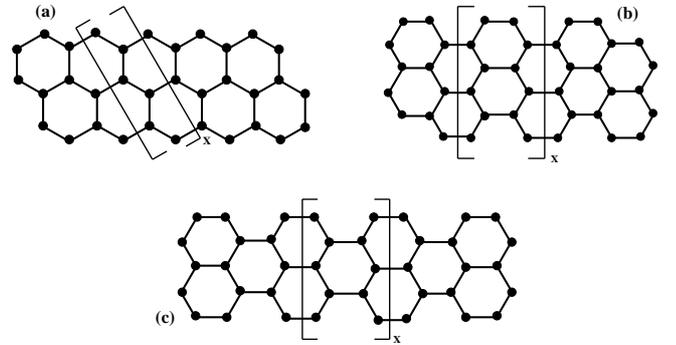}
\caption{\label{strucr} Molecular structures of (a) zigzag 3-ZGNR, (b) armchair 6-AGNR and (c) armchair 5-AGNR. 
The unit cell for each nanoribbon is indicated by the square brackets.} 
\end{center}
\end{figure}

The quasi one-dimensional character of the GNRs leads to confinement and enhanced electron repulsion between the 
$\pi$-electrons. Thus extended screening lengths and long range electron-electron interactions are expected in the semiconducting GNRs, 
and even in the metallic GNRs which are better described as zero gap semiconductors \cite{nakada,ezawa}.
Electron-electron interactions in GNRs and single-walled carbon nanotubes have been considered in the past within the
time-dependent density functional approach \cite{marinopoulos03,*marinopoulos04,tachikawa09,habenicht09} 
as well as the GW approximation accompanied by Bethe-Salpeter corrections \cite{spataru04,capaz05,yang07}.
These approaches take into account one electron-one hole (1e-1h) interactions, are equivalent to the single configuration 
interaction approximation of quantum chemistry \cite{Zhao04, Wang06, Wang07}, and have successfully predicted the excitonic character 
of the lowest optical absorption in the semiconductors. Our goal here, however, is to probe the consequences
of the strong four-fermion two electron-two hole (2e-2h) interactions. 
As in previous work \cite{Zhao04, Wang06, Wang07}, we probe the consequences of realistic electron-electron interactions 
on the optical and charge gaps in three narrow GNRs, the ZGNR with three carbon-carbon bonds across its width (3-ZGNR) and AGNRs with 
6 and 5 carbon-carbon bonds across their widths (6-AGNR and 5-AGNR), respectively (see {\bf Fig. \ref{strucr}}).
Beyond this, however, we also probe their ground state magnetic character and the spin gap.
We further determine the relative energy orderings of the lowest one- versus two-photon states
in all three nanoribbons. In the past, the experimental demonstration that the lowest two-photon state occurs
below the optical one-photon state in linear polyenes \cite{hudson-kohler-schulten}, in contradiction to the predictions of tight-binding
and Hartree-Fock theories \cite{hudson-kohler-schulten}, provided the most convincing demonstration of the strong electron correlations
in these systems. More recently, similar experimental results have also been obtained from nonlinear optical measurements of graphene
nanofragments \cite{Sandhu-Mazumdar}. Accurate computational investigation of the relative energy orderings of one- and two-photon
states also requires going beyond existing techniques. Finally, we compute nearest neighbor bond orders (nearest neighbor charge-transfers)
to examine the tendency to structural distortions.

For conjugated carbon systems, the Pariser-Parr-Pople (PPP) model, which assumes $\sigma-\pi$ separability and incorporates long-range 
electron-electron repulsions \cite{pppm,*pople}, is known to reproduce ground and excited state properties very well
\cite{schulten79-1,*schulten79-2,*schulten86,*schulten87,soos93,soos,baeriswyl}. It has also been demonstrated that the symmetrized  
DMRG method can provide highly accurate descriptions of ground and low-lying excited states within the PPP model \cite{raghu02-1,
*mano10,*simil13,*sukrit09,raghu02-2}. We report here symmetrized DMRG calculations on the three GNRs of {\bf Fig. \ref{strucr}}.

We briefly mention here existing related works about edge-magnetism in zigzag nanoribbons which was 
speculated quite early in the literature and has been studied rather extensively.
Fujita et al. have studied both armchair and zigzag graphene ribbons with tight-binding
approximation and reported the presence of a flat band and localized edge states near the Fermi level for zigzag
nanoribbons, resulting in high density of states near the Fermi level \cite{fujita}.
In armchair nanoribbons, these almost flat bands (which are a consequence of the topology of the $\pi$-conjugation) are absent.
In the correlated picture, these almost flat bands result in magnetic states whose spins are arranged
ferromagnetically at the edges \cite{fujita}. 
Opposite edges of the ribbon will have opposite alignment of spins making the ribbon non-magnetic in the thermodynamic limit.
Even in a general nanoribbon which cannot be classified as either perfectly armchair or perfectly zigzag, a few sequentially
placed zigzag sites can result in a significant density of states at the Fermi level resulting edge ferromagnetism \cite{nakada}.
Sasaki and coworkers studied graphene in the continuous model using the Weyl equation with a special
gauge field resulting from the local deformation in the $\pi$-backbone and confirmed the presence of localized states at zigzag 
edges \cite{sasaki06}. Consequences of the presence of these edge states in the quantum Hall effect
in graphene have also been studied within localized \cite{castroneto06} and continuum pictures \cite{levitov06}.
Wakabayashi et al. studied electronic and magnetic properties of GNRs in the presence of a magnetic field in the tight-binding 
approximation and proposed that zigzag nanoribbons will show diamagnetic behavior at high temperature and paramagnetic behavior at 
low temperature \cite{wakabayashi99}. Louie et al. have also showed that the edge-magnetic nature of ZGNRs can induce 
half-metallicity in the presence of a transverse electric field across the ribbon width, resulting in a spin current \cite{son06-2}.
The edge magnetism in zigzag nanoribbons has also been studied using the mean-field Hubbard model by Jung and coworkers \cite{jung}, 
employing the quantum Monte Carlo technique by Golor et al. \cite{golor} and by renormalization technique by Hikihara et al. 
\cite{hikihara} and experimentally at room temperature by Magda and coworkers \cite{magda}.
Instead of the presence of ferromagnetically aligned spins at the edges, all the above studies predicted a singlet ground state 
in ZGNR in the absence of an external field; however, a few density functional studies predicted a ferromagnetic ground state in ZGNR 
on doping \cite{okada01,*dutta08}. Dutta et al. have studied low-energy properties of both zigzag and armchair GNRs within the 
Hubbard model using quantum many-body configuration interaction method and predicted that the ground state of zigzag GNRs is a 
high-spin state while for armchair GNRs the ground state is a singlet \cite{datta2}. Spin density calculations at the edges 
of zigzag nanoribbons show the presence of both up and down spin at a given edge instead of predominance of a specific spin as
predicted by tight-binding and density functional theories. This study indicates that the picture can be significantly different 
in the presence of long-range electronic correlation.

There are also extensive studies of electronic properties such as band gaps and quasiparticle energies in GNRs.
Ezawa has studied bandgaps in a range of nanoribbons in the H\"{u}ckel picture and predicted the width 
dependence of the bandgaps in these systems \cite{ezawa}. He has also found that incorporation of edge effects by 
changing the transfer term or site energies of the edge sites has little effect on the band structure. 
Brey and Fertig studied the electronic structure of zigzag and armchair nanoribbons using the massless Dirac 
equation and their results are in agreement with the tight-binding results, except for narrow nanoribbons \cite{brey}.
They proposed that the continuum analysis of graphene can quantitatively predict the properties of these nanoribbons.
Louie et al. have computed the band gaps for zigzag and armchair GNRs by employing the first-principles approach within the local (spin) 
density approximation \cite{son06-1} and GW approximation with many-body Green's function technique \cite{yang07} and proposed 
analytical expressions for band gaps as a function of GNR widths. They argued all GNRs to be semiconducting, contradicting  
earlier tight-binding predictions. Recently, spin and charge gaps of the armchair polyacene have been studied within the PPP model
\cite{das}. It has been shown that the ground state of armchair polyacene is a singlet and the system is an insulator in the 
ground state.

This paper is organized as follows. In section II, we introduce the model Hamiltonian  
and briefly describe the DMRG scheme which we have employed in our study. In section III, 
we present and discuss our results for the three GNRs.
In the final section, we present a comparison of these three GNRs and summarize our results. 

\section{Theoretical model, the DMRG scheme for GNRs and symmetry subspaces}
\label{model}

\subsection{PPP Hamiltonian and parameters}

The PPP  Hamiltonian is written as, 

\begin{equation}
\begin{split} 
H = \sum_{<i,j>, \sigma}t_{ij}(a_{i\sigma}^\dagger a_{j\sigma}+&
a_{j\sigma}^\dagger a_{i\sigma}) + \frac{1}{2} \sum_{i}U_i n_i (n_i-1)\\
&+\sum_{i>j} V_{ij}(n_i-1)(n_j-1) ~~~
\label{hamlt}
\end{split}
\end{equation}

\noindent
where $<i,j>$ are the bonded pair of atoms and $t_{ij}$ is the corresponding hopping or transfer 
integral, $a_{i\sigma}^\dagger$ ($a_{i\sigma}$) is the creation (annihilation) 
operator at site i with spin $\sigma$ and $n_i$ is the number operator. U is the Hubbard on-site repulsion 
and $V_{ij}$ are the intersite electron-electron repulsion between carbon atoms i and j. $V_{ij}$ are 
obtained from the Ohno parametrization \cite{ohno,*klopman}, 

\begin{equation} 
V_{ij}=14.397 \left[\left(\frac{14.397}{U_i}\right)^2+r_{ij}^2\right]^{-\frac{1}{2}} 
\label{eqohno}
\end{equation}

\noindent
which is arrived at by interpolating between U at $r_{ij}=0$ and $e^2/r_{ij}$ for $r_{ij}\rightarrow$ $\infty$. 
In Eq. \ref{eqohno} the distances are in {\AA} while the energies are in eV ~\cite{race01,*race03}. We have taken the nearest neighbor distance 
between the carbon atoms as 1.42 {\AA} and we fixed parameters $t_{ij} = -2.40$ eV and $U=11.26$ eV as in many
of the previous studies involving carbon-based conjugated systems \cite{soos,raghu02-1,*mano10,*simil13,raghu02-2}; the U value chosen \cite{salem} is the sum of 
the ionization energy and the electron affinity of carbon and gives the energy change in the process $C C \rightarrow C^+ C^-$.

\subsection{The DMRG scheme}

We are interested in the properties of the three GNRs of 
{\bf Fig. \ref{strucr}} in the thermodynamic limit, which is reached from finite size scaling.
Exact studies of the model Hamiltonian are confined, at best, to about 
18 carbon sites and hence cannot be employed for extrapolation to the thermodynamic limit. Restricted 
configuration interaction approaches are not size consistent and cannot be employed with finite size 
scaling. The quantum Monte Carlo approach is also not suitable since we have long-range interactions in the model. 
For semiconducting narrow nanoribbons, the DMRG method is the method of choice as the area 
law of entanglement entropy holds and for the same DMRG cut-off similar accuracy is retained independent of the length 
of the nanoribbons \cite{scholl,hall}.
In the case of metallic nanoribbons, gapless low-lying excitations are present 
and the area law of entanglement entropy will not hold. This leads to increasing errors with increasing system 
sizes in the DMRG method if we employ a fixed cut-off in the number of block states $(M_l)$ for all system sizes. 
However, for finite systems in correlated models, there is always a finite gap in the excitation spectrum and 
by keeping a large number of block states, we can deduce correct excitation energies. 
While our calculations do not reproduce all behaviors of metals, we believe that the particular properties we are 
investigating can be obtained from extrapolations of these excitation energies. 
It has also been shown that for models with diagonal interactions in the real space such 
as the PPP model, the entanglement entropy is similar to those in the Hubbard and the Heisenberg models in 
one-dimension \cite{durga}. Taken together, the above justify the use of the DMRG approach for PPP calculations of GNRs. 

\begin{figure*}[tbp]
\begin{center} 
\includegraphics[width=15.0cm]{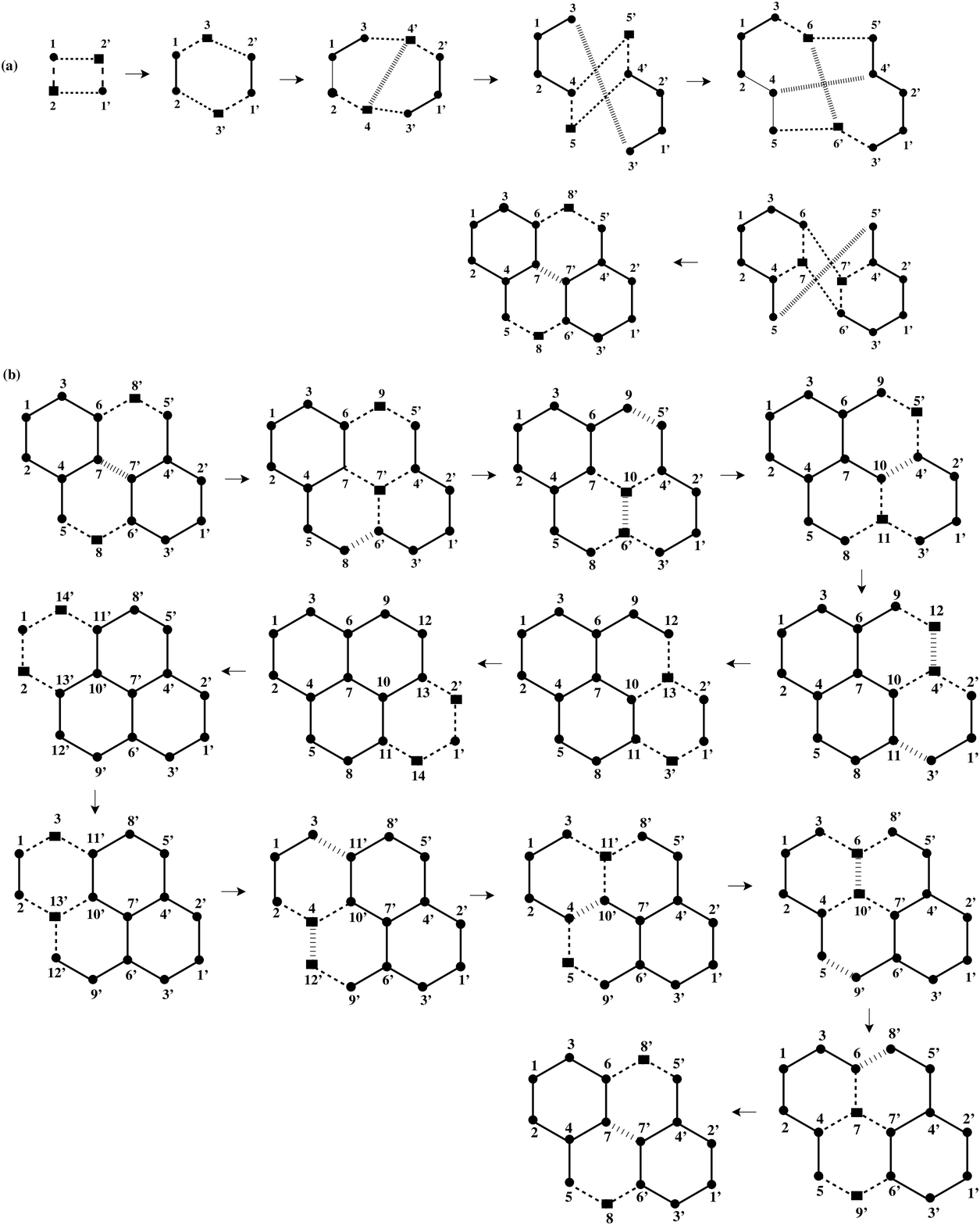}
\caption{\scriptsize{\footnotesize{\label{dmrg_zigzag} (a) Construction of 3-ZGNR of 16 sites in the infinite DMRG method 
starting from a small system (4 sites). The number of bonds between the new and the old sites at the intermediate system sizes are kept close to that in the 
final system for higher accuracy. At every step of the algorithm two new sites are added, one to the left sub-block (L) 
and other to the right sub-block (R). The sites in $L$ are denoted by unprimed numbers 
while those in $R$ are denoted by primed numbers. The newly added sites are shown by filled square ($\blacksquare$) while old sites 
are denoted by filled circles ($\bullet$). Solid lines are bonds within a sub-block. The broken lines denote the bonds between $\bullet$ and $\blacksquare$. 
Bonds between the two sub-blocks as well as the bond between $\blacksquare$ are denoted by hatched lines. 
(b) Scheme for sweeping in finite DMRG method for 16-site (2N=16) 3-ZGNR.
In the forward sweep, the left block size increases from $N-1$ to $2N-2$ sites as the right block size decreases from $N-1$ to 2 sites;
in the reverse sweep, the opposite happens.
During finite DMRG sweepings, the total system size remains constant. 
The corresponding figures for 6-AGNR and 5-AGNR are shown in {\bf Figs. $1$, $2$ and $3$} in the supplementary material \cite{supple}.  }}}
\end{center}
\end{figure*}

The DMRG method, discovered by White in 1992 divides a system block into two sub-blocks {\cite{white,*white-prb,scholl,hall,ramasesha}},
generally referred to as the left sub-block (L) and the right sub-block (R), while the wavefunction of the total system 
block is described in the direct product space of these two sub-block basis states. In the DMRG method, the Fock spaces 
of the two sub-blocks are approximated and it has been found that the best approximations of the sub-block Fock 
spaces can be obtained by retaining a small number of reduced density matrix eigenvectors corresponding to the highest reduced density matrix eigenvalues. 
The reduced density matrix of a chosen sub-block is obtained by treating the other block as an environment block and 
tracing the density over the states of the environment block. The reduced density matrix for a sub-block of size $l$, 
so obtained, is diagonalized and $M_l$ eigenvectors with the highest reduced density matrix eigenvalues are stored as 
column vectors of a $M_{l-1}d_{\sigma} \times M_{l}$ matrix where $M_{l-1}$ is the number of density matrix eigenstates retained
at the $l-1$-th iteration and $d_{\sigma}$ is the dimension of the Fock space of the new site added to the sub-block at the $l$-th iteration.
The Hamiltonian of the sub-blocks is renormalized using this $M_{l-1}d_{\sigma} \times M_l$ matrix and the matrices of 
site operators are also transformed to this new basis. All terms in the PPP Hamiltonian including 
the long-range correlation terms can be expressed using these renormalized site operators and the 
matrix elements of the system Hamiltonian can be obtained by taking appropriate direct products.
The next step in the DMRG algorithm involves expanding the system block by adding a few sites (usually two) to the previous 
system block. The Hamiltonian matrix is constructed in the direct product basis of the retained block states 
of the two sub-blocks and Fock states of the newly added sites. From the Hamiltonian matrix, desired 
eigenstates are obtained and the process of constructing the reduced density matrix, truncating the Fock space of
the augmented system and expanding the system by adding two additional sites and again solving for the desired 
eigenstate is repeated until we achieve the targeted system size. The dimension of the Hamiltonian 
matrix is independent of the system size as the number of block states retained to span the Fock 
space of the sub-blocks is fixed, independent of the physical size of the sub-block. The method described above is known as the infinite DMRG method. 

In order to obtain the behavior in the thermodynamic limit using only 
the infinite DMRG method, we would need to retain a rather large number of block states $(M_l)$ and 
go to much larger system sizes which is beyond our current computational capability. 
Instead, we have carried out finite DMRG calculations on systems of moderate sizes to obtain energy gaps with high accuracy and 
rely on finite size scaling to obtain the physical properties in the thermodynamic limit.

\begin{table}[tbp]
\begin{center}
\caption{\label{tab_huckel} Exact versus DMRG ground state energies (in eV) of 3-ZGNR, 6-AGNR and 5-AGNR 
within the non-interacting model (U=V$_{ij}$=0). The DMRG cut-off in the number of block states is 500.
}
\begin{ruledtabular}
\renewcommand{\arraystretch}{1.25}
\begin{tabular}{ccc} 
System type and size &\multicolumn{2}{c}{Ground state energy} \\
\cline{2-3}
& Exact calculation & DMRG method\\
\colrule 
3-ZGNR and 40 sites & -137.841  & -137.738  \\ 
6-AGNR and 40 sites & -139.503  & -139.312  \\  
5-AGNR and 40 sites & -137.859  & -137.813  \\ 
\end{tabular}
\end{ruledtabular}
\end{center}
\end{table}

The finite DMRG algorithm was introduced by White to improve the accuracy of finite system calculations.
In the infinite DMRG method, the density matrix of a p-site sub block is built from the eigenstates of a 
2p site system block. This leads to errors in the target system of 2N sites ($N>>p$). In order to correct this 
error, the p-site density matrices are iteratively constructed from the eigenstates of the 2N site
target system block, until convergence is achieved. This procedure is termed `sweeping' where iteratively the size 
of one of the sub-blocks increases at the expense of the other, while the total system block size remains unchanged. At the 
final step of one full sweep, sizes of the two sub-blocks become equal and the same as in the infinite DMRG step. 
The finite DMRG procedure for molecular systems is 
nontrivial but essential as the energies improve considerably following the sweeping. We have employed 
the finite DMRG algorithm with a block state cut-off $(M_l)$ of 500 and finite DMRG iteration of two sweeps
to compare the DMRG method results with the exact calculation results for all the three nanoribbons
within the non-interacting model (Table \ref{tab_huckel}). The ground state energies compare well with the exact 
nearest neighbor tight-binding energies in all 
cases with a cut-off in the number of block states $M_l=500$.  For the interacting models, we expect the DMRG method to be more accurate 
for the same cut-off because interacting systems are less entangled. The method of constructing the GNRs in the 
infinite DMRG procedure as well as the method of finite sweeps are shown in {\bf Fig. \ref{dmrg_zigzag}} and 
{\bf Figs. 1, 2 and 3} in the Supplementary Material \cite{supple}.

\begin{figure}[tpb]
\begin{center} 
\includegraphics[width=8.5cm,height=7.0cm]{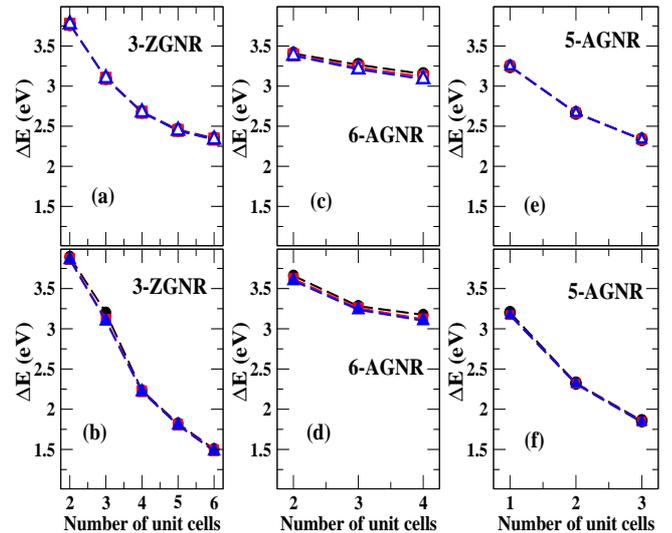}
\caption{\label{cutoff}{(Color online) Calculated lowest one-photon (optical) gaps ((a), (c) and (e)) and lowest two-photon gaps ((b), 
(d) and (f)) in short 3-ZGNR (left panel), 6-AGNR (middle panel) and 5-AGNR (right panel), plotted against the number of unit cells 
for different cut-off values of block states $(M_l)$. The symbols denote the following: black open circle-- one-photon gap with $M_l= \sim 750$;
red open square-- one-photon gap with $M_l= \sim 900$; blue open triangle-- one-photon gap with $M_l= \sim 1000$. Corresponding filled symbols denote 
two-photon gaps.}}
\end{center} 
\end{figure}

\subsection{Symmetry subspaces and one versus two-photon excitations}

The GNRs of interest possess $C_2$ symmetry along the axis perpendicular to the plane of 
the molecule, which we utilize in our computations as well as characterization of eigenstates. Eigenstates are labeled
$A$ or $B$ depending upon whether they are of even or odd parity with respect to $C_2$ operation. The 
PPP Hamiltonian conserves total spin $S$, but total spin conservation is difficult within the DMRG scheme with 
large cut-off in the block states. We exploit partial spin
symmetry by performing our calculations for the $S_z=0$ sector in which the Hamiltonian has spin inversion 
symmetry, corresponding to invariance of the Hamiltonian when all spins of the
system are reversed. This symmetry bifurcates the $S_z=0$ space into a subspace with even total spin, i.e., 
$S = 0, 2, 4,\cdot\cdot\cdot$ (hereafter designated as `e') and another with odd total spin, $S=1, 3, 5,\cdot\cdot\cdot$
(designated as `o'). Finally, the exactly half-filled band that we are investigating also exhibits charge-conjugation symmetry (CCS);
eigenstates are labeled even or odd (hereafter `+' or `-') depending upon the eigenvalue $\pm 1$ reached when operated by the
CCS operator \cite{pariser56,cizek74}.
The identity, the three symmetry operators and their products form an Abelian group
of eight elements. Hence, the $S_z=0$ sector gets subdivided into 8 subspaces. 

In general, the ground state is even with respect to all symmetry operations and lies in the ${}^eA^+$ subspace.
Optical one-photon states are reached by one application of the current operator $j$ on the ground state

\begin{equation}
\hat{j} = (i/\hbar) \sum_{<i,j>, \sigma}t_{ij}(a_{i\sigma}^\dagger a_{j\sigma}-a_{j\sigma}^\dagger a_{i\sigma})
\end{equation}

\noindent
which clearly changes the parity under $C_2$ symmetry while conserving $S_z$. It can also be shown that the application
of the $\hat{j}$ changes CCS \cite{pariser56}. Two-photon states are reached by one application of $\hat{j}$ on the optical state
(or two applications of the operator on the ground state), thus indicating that they also
lie in ${}^eA^+$ subspace. In the $S_z=1$ sector, spin inversion symmetry cannot be implemented and the 
lowest $S=1$ state is in the $B^+$ space. 
Using the symmetrized DMRG method ~\cite{pati} with a modified algorithm \cite{suryo}, all of these symmetries have been 
exploited in our calculations.

\begin{figure}[tpb]
\begin{center} 
\includegraphics[width=8.5cm,height=7.0cm]{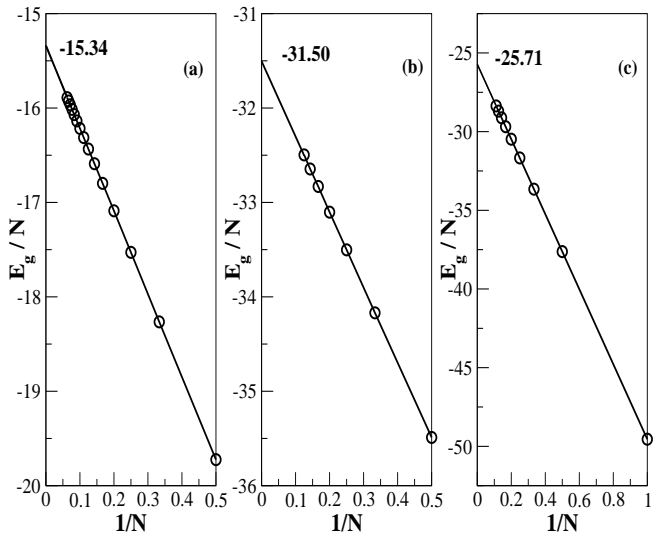}
\caption{\label{ebyn} Ground state energy in eV per unit cell versus 1/N for (a) 3-ZGNR, (b) 6-AGNR and (c) 5-AGNR  
within the PPP model. Here N is the number of unit cells as depicted in {\bf Fig. \ref{strucr}}.} 
\end{center} 
\end{figure}

In each symmetry subspace, we have calculated a few low-lying energy states of the Hamiltonian using Davidson's 
algorithm for symmetric sparse matrices. At each step of the DMRG algorithm, block states are computed from the average reduced density matrix
obtained from these eigenstates instead of the reduced density matrix of a single state. The average reduced density matrix is defined by 
$\rho=\sum_{k}\omega_{k}\rho_{k}$ where $\rho_k$ are the reduced density matrices corresponding to eigenstates $|k \rangle$ and $\omega_k$ 
are the weights of the corresponding eigenstates \cite{white-prb}.
We have taken $\omega_k=1/N_k$, where $N_k$ is the number of low-lying eigenstates computed in the symmetry subspace.
In what follows we define all energy gaps with respect to the ground state energy (thus the lowest one and two-photon gaps are the
energy differences between the corresponding eigenstates and the ground state).

In order to arrive at the desired cut-off in block states $M_l$ for our calculations, we have
calculated the lowest two-photon gaps and lowest optical gaps with different cut-offs for small systems of 3-ZGNR, 6-AGNR and 5-AGNR ({\bf Fig. \ref{cutoff}}). 
We note that $M_l \simeq 750$ is adequate for comparisons with experiments in all three GNRs.

\section{Results and Discussion}
\label{result}

We have used the unsymmetrized DMRG technique to calculate the ground state energies 
of these nanoribbons within the PPP model ({\bf Fig. \ref{ebyn}}). The excellent linear fit of the energies as 
a function of system size shows that the procedure is stable and accurate.  

\begin{figure}[tbp]
\begin{center} 
\includegraphics[width=8.5cm,height=7.0cm]{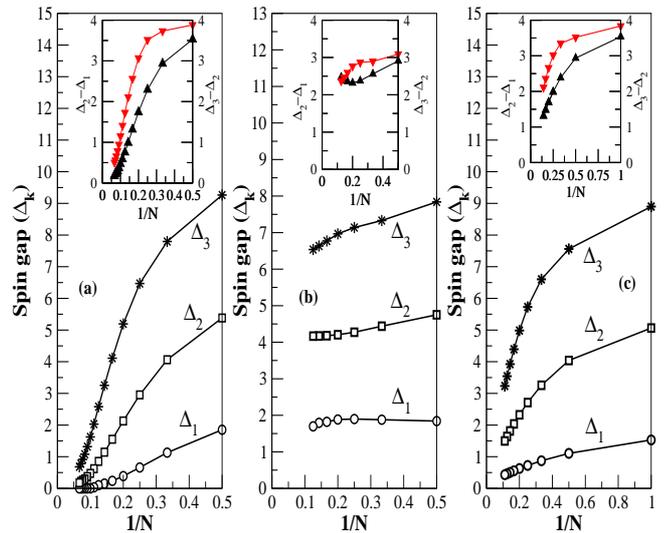}
\caption{\label{spingap_ribbons} (Color online) Spin gaps (see text) in eV versus the inverse of the number of unit cells for 
(a) 3-ZGNR (b) 6-AGNR and (c) 5-AGNR within the 
PPP model. Insets: $(\Delta_2-\Delta_1)$ [black solid triangle up] and $(\Delta_3-\Delta_2)$ [red solid triangle down], also versus 
the inverse of the number of unit cells.}
\end{center} 
\end{figure}
 
\subsection{\normalsize Spin gaps}

As mentioned above, it is difficult to exploit total S invariance in the DMRG scheme. 
We have therefore computed the lowest energy states in the different $S_z$ sectors; the total
$S$ are determined from the calculated energy gaps $\Delta_k=E_0(S_z=k)-E_0(S_z=0)$, where $E_0(S_z=k)$ is the 
lowest energy in the $S_z=k$ sector. In the absence of an external magnetic field, the different z-components of
a given total $S$ are degenerate; thus $\Delta_1=0$ implies that the ground state lies in the total spin $S=1$
subspace. This is true for arbitrary $\Delta_p$, and hence in general for $\Delta_1=\Delta_2=\cdot\cdot\cdot=\Delta_p=0$ and 
$\Delta_{p+1}>0$, the ground state has spin $S=p$.

We have shown the dependence of the computed energy gaps ($\Delta_k$) on the DMRG cut-off for a moderate-sized 
3-ZGNR system in Table. \ref{spin-cutoff}. We find that keeping $\sim750$ block states is sufficient for getting accurate gaps. 
Among all GNRs in this study, the energy gaps are smallest in 3-ZGNR; hence, we expect the same cut-off to be adequate for AGNRs also.

In Fig. \ref{spingap_ribbons}(a) we have plotted spin gaps $\Delta_k$ as a function of the 
inverse of the number of unit cells ($N$) for 3-ZGNR. For $N<14$, we find 
$\Delta_1>0$, indicating that the ground state is a singlet. For $N\geq14$, $S_z=1$ and $S_z=0$ states are degenerate 
(within the DMRG accuracy) which implies that the ground state has $S=1$. $\Delta_2>0$ in this region, but
becomes smaller as N is further increased.
It appears that S=2 will become the spin of the ground state for larger N values. Similarly, the gap between $S_z=3$ 
and $S_z=0$ states also decreases with increasing N. The inset in Fig. \ref{spingap_ribbons}(a) shows the behavior
of $\Delta_3-\Delta_2$ and $\Delta_2-\Delta_1$, both of which rapidly approach zero at large $N$. 
From these trends in the the spin gaps, we predict that the ground state
of this system is ferromagnetic in the thermodynamic limit.

\begin{table}[tbp]
\begin{center}
\caption{\label{spin-cutoff}Different $\Delta_k$ values in eV calculated for 3-ZGNR with 5 unit cells, for different number of retained block states. 
The change in the energy gaps for the different cut-off values are not significant.}
\begin{ruledtabular}
\renewcommand{\arraystretch}{1.25}
\begin{tabular}{cccc} 
Block states cut-off & $\Delta_1$ & $\Delta_2$ & $\Delta_3$ \\
\colrule 
700 & 0.290 & 1.608 & 4.174   \\ 
750 & 0.292 & 1.610 & 4.177   \\ 
800 & 0.293 & 1.612 & 4.180   \\ 
\end{tabular}
\end{ruledtabular}
\end{center}
\end{table}

The spin gaps $\Delta_k$ for 6-AGNR are shown in {\bf Fig. \ref{spingap_ribbons}(b)}. 
$\Delta_1$ now exhibits  weak size dependence and continues to be nonzero even in the thermodynamic limit.   
The extrapolated $\Delta_1$ in the thermodynamic limit is 1.43 eV which is actually larger than 
the non-interacting tight-binding bandgap of  1.19 eV. This implies that the ground state of this system is a singlet. 
Indeed $\Delta_1$ is larger than that in polyacene which has a spin gap of $\sim$0.5 eV in the thermodynamic limit \cite{raghu02-2}. 
The spin gaps $\Delta_2$ and $\Delta_3$ are also large and weakly size-dependent, as are the differences between the spin gaps
$\Delta_2-\Delta_1$ and $\Delta_3-\Delta_2$, plotted in the inset of {\bf Fig. \ref{spingap_ribbons}(b).}
In general $E(S)>E(S^{\prime})$ for $S>S^{\prime}$ here, where $E(S)$ is the lowest energy in the total spin $S$ subspace, and the
corresponding energy differences are large.

For 5-AGNR, the spin gaps between different $S_z$ sectors and $S_z=0$ are shown in {\bf Fig. \ref{spingap_ribbons}(c)}.
The extrapolated ($\Delta_1$) in the thermodynamic limit is 0.15 eV, indicating that the ground state is spin singlet.
From the N-dependent behavior of $\Delta_2$ and $\Delta_3$ (see in particular
the inset) it is conceivable that the energy spectrum above $\Delta_1$ may be gapless.

\begin{figure}[tbp]
\begin{center}
\includegraphics[width=8.0cm]{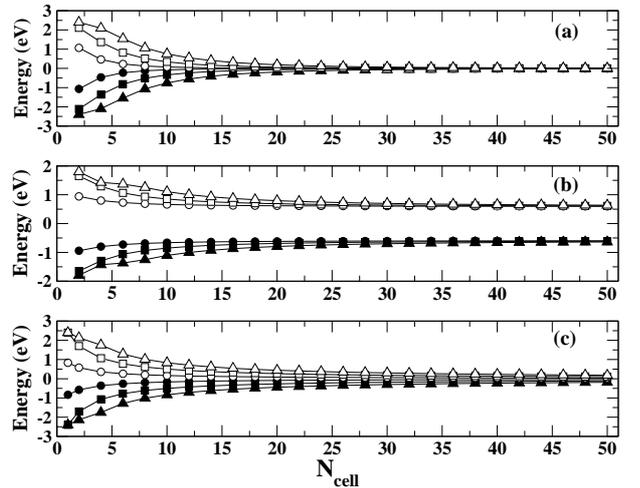}
\caption{\label{huckel} A few energy levels near the Fermi level for (a) 3-ZGNR (b) 6-AGNR and 
(c) 5-AGNR within the tight-binding model are plotted as a function of number of unit cells. The hopping energy 
considered is 2.40 eV. The symbol index is same for all the three panels and are as follows: HOMO energy level (solid circle); 
HOMO--1 energy level (solid square); HOMO--2 energy level (solid triangle up); LUMO energy level (open circle); LUMO+1 energy 
level (open square) and LUMO+2 energy level (open triangle up).}
\end{center}
\end{figure}

It is instructive to see what is to be expected for the spin gaps in the non-interacting limit. The energy levels of the
frontier molecular orbitals of the three GNRs are shown in Fig. \ref{huckel}. We see that in 3-ZGNR, the energy gap between frontier 
orbitals approaches zero rapidly implying that switching on exchange interaction will lead to a high spin ground state. In 6-AGNR, the gap 
between bonding and antibonding frontier orbitals is finite for all system sizes. Introduction of electron-electron interaction will therefore 
not change the spin of the ground state and the ground state will always be a singlet. In 5-AGNR, the gap between the bonding and antibonding 
frontier orbitals progressively decreases but remain finite for large system sizes. The small band gap implies a small spin gap in the interacting picture.

In summary, the effects of electron-electron interactions on the three GNRs we have studied are very different
and could not have been anticipated from their tight-binding electronic structures. The ferromagnetic ground state in
3-ZGNR is different from the edge-state ferromagnetism found earlier in wider ZGNRs \cite{fujita,son06-1,son06-2}. More interestingly, while within
tight-binding theory 5-ZGNR is metallic and is hence expected to be without a spin gap, we find a small but nonzero spin gap here,
although the spectrum of spin excitations above the lowest gap may be gapless. As we show in section~\ref{result}(C), this system also 
has a nonzero exciton binding energy for nonzero electron-electron interactions. 
In 6-AGNR, which is a band semiconductor, both charge and spin gaps are expected within the tight-binding model. 
Our calculated results indicate that electron-electron interactions further enhance the spin gap here. We discuss these effects further
in section section~\ref{result}(B).

\begin{figure}[tbp]
\begin{center} 
\includegraphics[width=8.0cm]{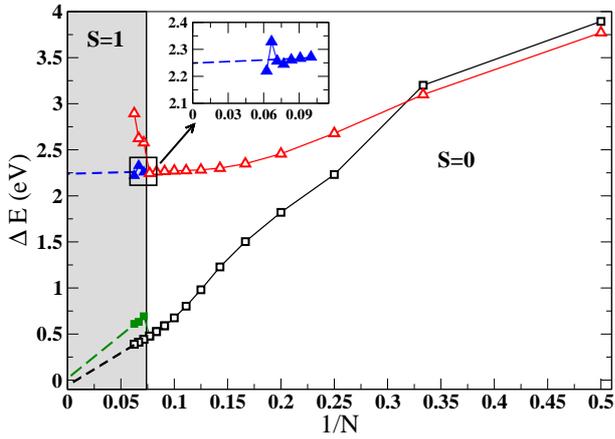}
\caption{\label{zigzag_opg} (Color online) Lowest optical and two-photon gaps in 3-ZGNR versus 
the inverse of number of unit cells.  As the ground state spin value changes on increasing the number of 
monomer units, the lowest optical gaps and lowest two-photon gaps in both $S=0$ and $S=1$ sectors are plotted. Symbols represent the
following: lowest optical gap in singlet space ({\color{Red}$\mathbf{\vartriangle}$}); lowest two-photon gap in singlet space ($\square$);
lowest optical gap in triplet space ({\color{Blue}$\blacktriangle$}); lowest two-photon gap in triplet space ({\color{ForestGreen}$\blacksquare$}). 
Inset: Magnified plot of the lowest optical gaps and its extrapolation in 3-ZGNR systems, 11-16 monomer units.}
\end{center} 
\end{figure}

\subsection{\normalsize Excited state ordering of one- versus two-photon states}

As already mentioned in section~\ref{model}(C), the occurrence of the lowest two-photon states below the
lowest optical one-photon states in linear polyenes \cite{Hudson-Kohler,Schulten-Karplus} was the strongest evidence for higher order Coulomb interaction
effects beyond 1e-1h interactions. It is therefore of interest to determine the excited state ordering in these narrow GNRs
we are probing; this is particularly so because should the present systems become available experimentally, 
the corresponding two-photon states can be reached by a variety of nonlinear spectroscopic
techniques, and our theoretical predictions tested.

We have obtained the low-lying one- and two-photon excited states for all three GNRs  
within the PPP model. We have done calculations starting from 2 units up to a maximum of 16 units of 3-ZGNR, 
from 2 units up to a maximum of 8 units of 6-AGNR and from 1 unit up to 9 units of 5-AGNR, which correspond to 
about 100 carbon atoms in the largest systems studied.
For extrapolation of different energy gaps in all three nanoribbons, we have considered the largest system sizes, specifically,
from $N=10$ to $N=16$ for 3-ZGNR, from $N=5$ to $N=8$ in 6-AGNR and from $N=4$ to $N=9$ in 5-AGNR ($N$ is the number of unit cells). 

\begin{figure}[btp]
\begin{center} 
\includegraphics[width=8.0cm]{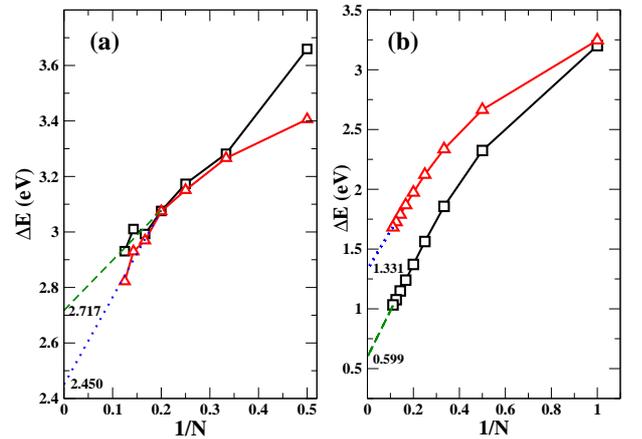}
\caption{\label{tpg_opg} (Color online) Lowest one-photon (optical) ({\color{Red}$\vartriangle$}) and two-photon ($\square$) gaps  
in (a) 6-AGNR and (b) 5-AGNR, versus the inverse of the number of unit cells. The ground state remains 
in both the cases in $S=0$ space. The extrapolated one-photon and two-photon gaps are also indicated in the figures.}
\end{center} 
\end{figure}

In 3-ZGNR, the lowest two-photon state occurs above the lowest optical state 
for system sizes up to three units, but this energy ordering is reversed in larger systems ({\bf Fig. \ref{zigzag_opg}})
Similar size-dependence has also been observed in linear polyenes and is expected from theoretical considerations \cite{hudson-kohler-schulten}.
As pointed out above, the ground state of 3-ZGNR changes beyond a certain size. We have therefore plotted in {\bf Fig. \ref{zigzag_opg}} 
the lowest optical and two-photon gaps for both $S=0$ and $S=1$ ground states. 
In the {\bf Fig. \ref{zigzag_opg}} inset, the data points are shown on an expanded scale and appear to be scattered. However, 
the calculated standard deviation {\cite{chapra}} for the linear fit is small ($0.036$ eV).
The extrapolated value of the optical gap in the thermodynamic limit is found to be 2.25 eV, irrespective 
of the ground state spin, which is in contradiction to the prediction of a metallic state for this nanoribbon 
within one-electron theory (see also section~\ref{intro}). Note that for system sizes where the ground state is a triplet, 
the singlet optical gap increases with system size. This is an artifact that is irrelevant for the real system with a magnetic ground 
state. As shown in the figure as well as in the inset, the triplet and singlet optical gaps lie 
on the same continuous curve if the data from the artificially large singlet gaps are ignored.
However, the two-photon gaps in both singlet and triplet spaces extrapolate to zero in the thermodynamic limit.
In linear polyenes, the lowest two-photon state is known to be a quantum-entangled
state of two triplets with overall spin angular momentum of zero. The zero gap two-photon state in the present case is
to be anticipated, should the same theoretical description as a triplet-triplet state persist here.

In 6-AGNR, the lowest optical state always remains below the lowest two-photon state for all system sizes. 
The extrapolated one- and two-photon gaps at the thermodynamic limit are quite distinct ({\bf Fig \ref{tpg_opg}a}).
The extrapolated value of the optical gap is $2.45$ eV while that of the two-photon gap is $2.72$ eV. 
The standard deviations of the linear fits of the 
optical gaps and the two-photon gaps are $0.032$ eV and $0.026$ eV respectively.
The larger two-photon gap here is a reflection of the predominantly band semiconductor character of 6-AGNR, therefore the 
one-electron contribution to the optical gap must be larger than the many-electron contribution. The relative 
energy locations of the optical and two-photon states in 6-AGNR suggest that these systems will be fluorescent.  
In the case of 5-AGNR, on the other hand, the optical state always remains above the lowest two-photon state, although 
the lowest two-photon gap does not vanish in the thermodynamic limit ({\bf Fig \ref{tpg_opg}b}) but saturates at a value 
of $0.60$ eV. The optical gap in 5-AGNR extrapolates to 1.33 eV at the thermodynamic limit. An optical gap larger than the 
two-photon gap in 5-AGNR is a consequence of the former being dominated by Coulomb as opposed to band contribution.

\begin{figure}[tbp]
\begin{center} 
\includegraphics[width=8.0cm]{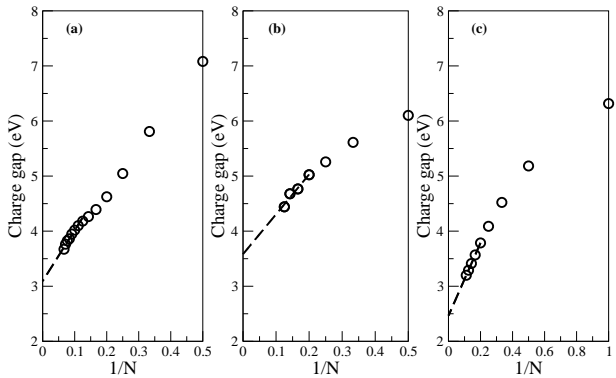}
\caption{\label{chrgegap} Extrapolation of the charge gaps against the
inverse of the number of unit cells, for 
(a) 3-ZGNR, (b) 6-AGNR and (c) 5-AGNR, respectively.} 
\end{center} 
\end{figure}

\subsection{\normalsize Charge gap}

We have calculated the charge gaps in these nanoribbons to explore the conducting nature in the thermodynamic limit, 
in the presence of long-range interactions ({\bf Fig. \ref{chrgegap}}). The charge gap of a system with N unit cells, $\Delta_c(N)$ is defined as the 
energy required to create a well separated electron-hole pair from the ground state of the system: 
$\Delta_c (N) =E^+(N)+E^-(N)-2E^0(N)$, where $E^+(N)$ and $E^-(N)$ are the ground state energies of the cation and 
anion respectively and $E^0(N)$ is the ground state energy of the neutral system.  
In the thermodynamic limit, the charge gaps of 3-ZGNR, 6-AGNR and 5-AGNR  
extrapolate to 3.09 eV, 3.58 eV and 2.46 eV respectively. Interestingly, the exciton binding energy ($E_{xb}$) of the optical state ($E_{1ph}$) 
which is measured as $E_{xb}=E_{cg}-E_{1ph}$ is quite small in all the three GNRs compared to other known 
organic conjugated systems. The exciton binding energies are 0.84 eV in 3-ZGNR, 
1.13 eV in 6-AGNR and 1.13 eV in 5-AGNR. Hence,
these conjugated systems can have importance in molecular photovoltaics as the electron-hole pair 
can be relatively easily disassociated.

\begin{table*}[htbp]
\begin{center}
\caption{\label{tab_bndord1} Bond orders of an interior unit of 3-ZGNR. Bond orders in systems with $\le 13$ 
monomer units are given in the first row while those for systems with $>13$ monomer units are given in the second row.} 
\begin{ruledtabular}
\begin{tabular}{cccccccccccc} 
State      &  1   &  2   &   3  &   4  &  5   &  6   &  7   &  8   &  9   &  10  &  11   \\ 
\colrule
Ground state & 0.54 & 0.56 & 0.47 & 0.47 & 0.52 & 0.52 & 0.52 & 0.47 & 0.47 & 0.56 & 0.54  \\
             & 0.56 & 0.54 & 0.47 & 0.47 & 0.52 & 0.53 & 0.52 & 0.47 & 0.47 & 0.54 & 0.56  \\ 
\colrule
Optical state & 0.54 & 0.57 & 0.46 & 0.46 & 0.54 & 0.50 & 0.54 & 0.46 & 0.46 & 0.57 & 0.54  \\
              & 0.56 & 0.54 & 0.46 & 0.46 & 0.51 & 0.54 & 0.51 & 0.46 & 0.46 & 0.54 & 0.56   \\ 
\colrule
Two-photon state & 0.52 & 0.58 & 0.48 & 0.47 & 0.50 & 0.54 & 0.50 & 0.47 & 0.48 & 0.58 & 0.52  \\
                 & 0.54 & 0.56 & 0.47 & 0.47 & 0.51 & 0.52 & 0.51 & 0.47 & 0.47 & 0.56 & 0.55   \\ 
\colrule
Spin state & 0.56 & 0.53 & 0.47 & 0.47 & 0.52 & 0.53 & 0.52 & 0.47 & 0.47 & 0.53 & 0.57  \\
           & 0.54 & 0.55 & 0.47 & 0.47 & 0.53 & 0.52 & 0.53 & 0.47 & 0.47 & 0.55 & 0.54  \\ 
\end{tabular}
\end{ruledtabular}
\end{center}
\end{table*}

\begin{table*}[htbp]
\begin{center}
\caption{\label{tab_bndord2} Bond orders of an interior unit of 6-AGNR.}
\begin{ruledtabular}
\begin{tabular}{cccccccccccccc}
State      &  1   &  2   &   3  &   4  &  5   &  6   &  7   &  8   &  9   &  10  &  11  &  12  &  13  \\ 
\colrule
Ground state & 0.58 & 0.59 & 0.37 & 0.57 & 0.56 & 0.44 & 0.45 & 0.51 & 0.57 & 0.56 & 0.58 & 0.58 & 0.66 \\
\colrule
Optical state & 0.49 & 0.50 & 0.48 & 0.53 & 0.50 & 0.48 & 0.50 & 0.51 & 0.53 & 0.50 & 0.49 & 0.50 & 0.72 \\
\colrule
Two-photon state & 0.57 & 0.57 & 0.38 & 0.54 & 0.54 & 0.47 & 0.47 & 0.50 & 0.55 & 0.54 & 0.56 & 0.58 & 0.65 \\ 
\colrule
Triplet state & 0.48 & 0.48 & 0.50 & 0.52 & 0.48 & 0.48 & 0.51 & 0.51 & 0.52 & 0.48 & 0.47 & 0.48 & 0.73 \\
\end{tabular}
\end{ruledtabular}
\end{center}
\end{table*}

\begin{table*}[!]
\begin{center}
\caption{\label{tab_bndord3} Bond orders of an interior unit of 5-AGNR.}
\begin{ruledtabular}
\begin{tabular}{cccccccccccccc} 
State      &  1   &  2   &   3  &   4  &  5   &  6   &  7   &  8   &  9   &  10  &  11  &  12  &  13  \\ 
\colrule
Ground state & 0.60 & 0.39 & 0.60 & 0.63 & 0.51 & 0.51 & 0.52 & 0.51 & 0.51 & 0.61 & 0.39 & 0.60 & 0.63 \\
\colrule
Optical state & 0.57 & 0.42 & 0.57 & 0.67 & 0.51 & 0.51 & 0.51 & 0.51 & 0.52 & 0.57 & 0.42 & 0.57 & 0.67 \\
\colrule
Two-photon state & 0.50 & 0.48 & 0.51 & 0.72 & 0.52 & 0.52 & 0.49 & 0.52 & 0.52 & 0.51 & 0.48 & 0.50 & 0.71 \\ 
\colrule
Triplet state & 0.52 & 0.47 & 0.51 & 0.72 & 0.52 & 0.52 & 0.50 & 0.52 & 0.52 & 0.52 & 0.47 & 0.51 & 0.72 \\
\end{tabular}
\end{ruledtabular}
\end{center}
\end{table*}

\subsection{\normalsize Bond order}

\begin{figure}[t]
\begin{center} 
\includegraphics[width=7.0cm]{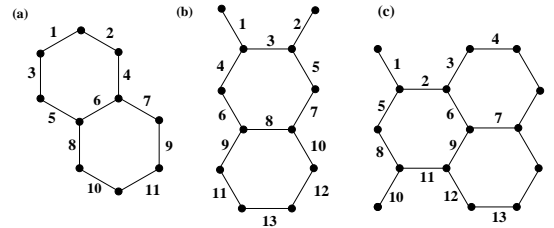}
\caption{\label{bndindx} Bond indices for the interior bonds of (a) 3-ZGNR, (b) 6-AGNR and (c) 5-AGNR.} 
\end{center} 
\end{figure}

We have calculated bond orders for the ground state and a few low-lying excited states
of the largest systems we studied. The bond order ($p_{ij}$) for the $<i,j>$ bond in a given state 
$|R\rangle$ is defined as  
$(-\frac{1}{2}) \sum_{\sigma}\langle R|(a_{i\sigma}^\dagger a_{j\sigma}+h.c.)|R \rangle$, and their deviation 
from an average value shows the tendency for the bond to distort. If the bond order is more (less) 
than the average, we expect the bond to shorten (lengthen) at equilibrium geometry. 
The numbering of bonds in the interior units of 3-ZGNR, 6-AGNR and 5-AGNR  
are shown in {\bf Fig. \ref{bndindx}}. The bond orders towards the ends of the ribbons are 
normally different from the bond orders in the interior because of edge effects. 
In Table \ref{tab_bndord1}, the bond orders of the interior unit of 3-ZGNR  
are given. In 3-ZGNR series, we have given bond orders for two different system sizes; 
one with 13 monomer units, which has a singlet ground state and another one with 16 monomer units 
with a triplet ground state. In both systems, the bond order differences between edge bonds in the ground state 
are small, implying almost uniform geometry of the edges. The rung bond orders are slightly smaller than the edge bonds, 
implying longer rung bonds than edge bonds at equilibrium geometry. The ground state,
the optical state and the lowest spin state all have similar geometries. The bond orders in the singlet and
triplet two-photon states are similar away from middle of the ribbon. However, in the middle of the ribbon,
the triplet two-photon state has marked single and double bond character in the top and bottom edge bonds
(bonds 1, 2, 10 and 11).

The bond orders in 6-AGNR systems show somewhat different behavior. 
In the ground state, bonds on the ring and on the exterior of the edge (bond 13) show 
tendency to contract, while bond 3 on the edge which is on the interior shows a tendency to expand. 
There appears to be a period (short-short-short-long) on one 
edge and out of phase by two bonds on the opposite edge. In the two-photon state, the bond orders remain 
almost the same as those in the ground state. In the one-photon 
state, there is considerable deviation in the bond orders in the central region of the nanoribbon, but the 
effect decreases towards the ends. However in the triplet state it appears that this distortion 
is much more pronounced. Thus, the equilibrium geometries of the excited states are quite different from 
that in the ground state and we may expect larger Stokes shifts in the spectra of 6-AGNR
compared to 3-ZGNR.

The bond orders on the edges of the 5-AGNR are similar to those in the 
6-AGNR. Except for the bonds which have rather large and small bond order values, 
all other bonds have nearly equal bond orders. So the distortion is expected more on the edges than 
in the interior. In all excited states, the interior bonds remain almost unperturbed. The edge bond 
structure in the central region of the nanoribbon shifts from short-long-short-short modulation to 
long-long-long-short modulation, while being unchanged towards the ends of the nanoribbon. 
The extent of distortion in the central region is highest in the two-photon and triplet states as compared 
to the one-photon state.

\section{Conclusion}

We have studied correlated electronic properties of 3-ZGNR, 6-AGNR and 5-AGNR  
within the PPP Hamiltonian with long range Coulomb interactions. 
In all three cases the ground state, as well as excited state behaviors are qualitatively different from
the predictions of tight-binding theory, as summarized below.

We find 3-ZGNR to be a magnetic semiconductor with a Mott-Hubbard optical gap and a substantive exciton binding energy. The
lowest two-photon state is gapless. The semiconducting behavior of 3-ZGNR is not entirely unanticipated, as
similar Mott-Hubbard semiconducting behavior was previously predicted from correlated-electron calculations {\cite{louie97,balents97}} 
and also experimentally demonstrated {\cite{deshpande09}} in narrow single-walled carbon nanotubes which would be metallic within one-electron theory.
As with 3-ZGNR, 5-AGNR is expected to exhibit metallic behavior within tight-binding theory,
but it is also found to be a Mott-Hubbard semiconductor here, with now however a singlet ground state with small spin gap. The two-photon state
is again below the optical gap. 5-AGNR thus resembles an idealized trans-polyacetylene strand, where bond-dimerization 
leads to a spin gap, and where the occurrence of the two-photon state below the one-photon optical gap is believed to be a signature  
of greater Mott-Hubbard contribution to the optical gap than the contribution due to the Peierls bond-dimerization \cite{baeriswyl}.
In 6-AGNR, the one-electron optical gap is enhanced by Coulomb interactions, there occurs a large spin gap and the two-photon state occurs
above the optical gap. The overall behavior is reminiscent now of the conjugated polymers poly-paraphenylene and poly-paraphenylenevinylene, 
where the optical gap is dominated by the one-electron gap expected in systems with unit cells containing an even number of carbon atoms \cite{soos93}. 
The three GNRs we have studied thus span the full range of behavior expected in quasi-one dimensional correlated-electron systems. Conversely,
the apparent similarities between these narrow GNRs and conjugated polymers reflect the deep and fundamental universality that exists
among low-dimensional correlated-electron systems. Experimental tests of our theoretical predictions will provide
fresh insight on the role of electron-electron interactions in carbon nanostructures. It is also of interest to determine how these features evolve upon controlled
increase in the widths of GNRs. This is a topic of future research.

\begin{acknowledgments}

We acknowledge Indo-US Science and Technology Forum for financial assistance through a Joint Center, that made this collaboration
possible. SR thanks DST for providing additional 
financial assistance. SM acknowledges partial support from National Science Foundation grant DMR-1151475 and the UA-REN Faculty Exploratory Research Grant.
SP acknowledges CSIR India for senior research fellowship. 

\end{acknowledgments}

\bibliography{manuscript}

\begin{thebibliography}{81}%
\makeatletter
\providecommand \@ifxundefined [1]{%
 \@ifx{#1\undefined}
}%
\providecommand \@ifnum [1]{%
 \ifnum #1\expandafter \@firstoftwo
 \else \expandafter \@secondoftwo
 \fi
}%
\providecommand \@ifx [1]{%
 \ifx #1\expandafter \@firstoftwo
 \else \expandafter \@secondoftwo
 \fi
}%
\providecommand \natexlab [1]{#1}%
\providecommand \enquote  [1]{``#1''}%
\providecommand \bibnamefont  [1]{#1}%
\providecommand \bibfnamefont [1]{#1}%
\providecommand \citenamefont [1]{#1}%
\providecommand \href@noop [0]{\@secondoftwo}%
\providecommand \href [0]{\begingroup \@sanitize@url \@href}%
\providecommand \@href[1]{\@@startlink{#1}\@@href}%
\providecommand \@@href[1]{\endgroup#1\@@endlink}%
\providecommand \@sanitize@url [0]{\catcode `\\12\catcode `\$12\catcode
  `\&12\catcode `\#12\catcode `\^12\catcode `\_12\catcode `\%12\relax}%
\providecommand \@@startlink[1]{}%
\providecommand \@@endlink[0]{}%
\providecommand \url  [0]{\begingroup\@sanitize@url \@url }%
\providecommand \@url [1]{\endgroup\@href {#1}{\urlprefix }}%
\providecommand \urlprefix  [0]{URL }%
\providecommand \Eprint [0]{\href }%
\providecommand \doibase [0]{http://dx.doi.org/}%
\providecommand \selectlanguage [0]{\@gobble}%
\providecommand \bibinfo  [0]{\@secondoftwo}%
\providecommand \bibfield  [0]{\@secondoftwo}%
\providecommand \translation [1]{[#1]}%
\providecommand \BibitemOpen [0]{}%
\providecommand \bibitemStop [0]{}%
\providecommand \bibitemNoStop [0]{.\EOS\space}%
\providecommand \EOS [0]{\spacefactor3000\relax}%
\providecommand \BibitemShut  [1]{\csname bibitem#1\endcsname}%
\let\auto@bib@innerbib\@empty
%</preamble>
\bibitem [{\citenamefont {Saito}\ \emph {et~al.}(1998)\citenamefont {Saito},
  \citenamefont {Dresselhaus},\ and\ \citenamefont {Dresselhaus}}]{saito}%
  \BibitemOpen
  \bibinfo {editor} {\bibfnamefont {R.}~\bibnamefont {Saito}}, \bibinfo
  {editor} {\bibfnamefont {G.}~\bibnamefont {Dresselhaus}}, \ and\ \bibinfo
  {editor} {\bibfnamefont {M.~S.}\ \bibnamefont {Dresselhaus}},\ eds.,\
  \href@noop {} {\emph {\bibinfo {title} {Physical Properties of Carbon
  Nanotubes}}}\ (\bibinfo  {publisher} {Imperial College Press},\ \bibinfo
  {address} {London},\ \bibinfo {year} {1998})\BibitemShut {NoStop}%
\bibitem [{\citenamefont {Dresselhaus}\ \emph {et~al.}(1988)\citenamefont
  {Dresselhaus}, \citenamefont {Dresselhaus}, \citenamefont {Sugihara},
  \citenamefont {Spain},\ and\ \citenamefont {Goldberg}}]{dresselhaus}%
  \BibitemOpen
  \bibinfo {editor} {\bibfnamefont {M.~S.}\ \bibnamefont {Dresselhaus}},
  \bibinfo {editor} {\bibfnamefont {G.}~\bibnamefont {Dresselhaus}}, \bibinfo
  {editor} {\bibfnamefont {K.}~\bibnamefont {Sugihara}}, \bibinfo {editor}
  {\bibfnamefont {I.~L.}\ \bibnamefont {Spain}}, \ and\ \bibinfo {editor}
  {\bibfnamefont {H.~A.}\ \bibnamefont {Goldberg}},\ eds.,\ \href@noop {}
  {\emph {\bibinfo {title} {Graphite Fibers and Filaments}}}\ (\bibinfo
  {publisher} {Springer-Verlag},\ \bibinfo {address} {Berlin},\ \bibinfo {year}
  {1988})\BibitemShut {NoStop}%
\bibitem [{\citenamefont {Geim}\ and\ \citenamefont {Novoselov}(2007)}]{geim}%
  \BibitemOpen
  \bibfield  {author} {\bibinfo {author} {\bibfnamefont {A.~K.}\ \bibnamefont
  {Geim}}\ and\ \bibinfo {author} {\bibfnamefont {K.~S.}\ \bibnamefont
  {Novoselov}},\ }\href@noop {} {\bibfield  {journal} {\bibinfo  {journal}
  {Nature Materials}\ }\textbf {\bibinfo {volume} {6}},\ \bibinfo {pages} {183}
  (\bibinfo {year} {2007})}\BibitemShut {NoStop}%
\bibitem [{\citenamefont {Neto}\ \emph {et~al.}(2009)\citenamefont {Neto},
  \citenamefont {Guinea}, \citenamefont {Peres}, \citenamefont {Novoselov},\
  and\ \citenamefont {Geim}}]{neto}%
  \BibitemOpen
  \bibfield  {author} {\bibinfo {author} {\bibfnamefont {A.~H.~C.}\
  \bibnamefont {Neto}}, \bibinfo {author} {\bibfnamefont {F.}~\bibnamefont
  {Guinea}}, \bibinfo {author} {\bibfnamefont {N.~M.~R.}\ \bibnamefont
  {Peres}}, \bibinfo {author} {\bibfnamefont {K.~S.}\ \bibnamefont
  {Novoselov}}, \ and\ \bibinfo {author} {\bibfnamefont {A.~K.}\ \bibnamefont
  {Geim}},\ }\href@noop {} {\bibfield  {journal} {\bibinfo  {journal} {Rev.
  Mod. Phys.}\ }\textbf {\bibinfo {volume} {81}},\ \bibinfo {pages} {109}
  (\bibinfo {year} {2009})}\BibitemShut {NoStop}%
\bibitem [{\citenamefont {Allena}\ \emph {et~al.}(2010)\citenamefont {Allena},
  \citenamefont {Tung},\ and\ \citenamefont {Kaner}}]{allen}%
  \BibitemOpen
  \bibfield  {author} {\bibinfo {author} {\bibfnamefont {M.~J.}\ \bibnamefont
  {Allena}}, \bibinfo {author} {\bibfnamefont {V.~C.}\ \bibnamefont {Tung}}, \
  and\ \bibinfo {author} {\bibfnamefont {R.~B.}\ \bibnamefont {Kaner}},\
  }\href@noop {} {\bibfield  {journal} {\bibinfo  {journal} {Chem. Rev.}\
  }\textbf {\bibinfo {volume} {110}},\ \bibinfo {pages} {132} (\bibinfo {year}
  {2010})}\BibitemShut {NoStop}%
\bibitem [{\citenamefont {{Das Sarma}}\ \emph {et~al.}(2011)\citenamefont {{Das
  Sarma}}, \citenamefont {Adam}, \citenamefont {Hwang},\ and\ \citenamefont
  {Rossi}}]{sarma}%
  \BibitemOpen
  \bibfield  {author} {\bibinfo {author} {\bibfnamefont {S.}~\bibnamefont {{Das
  Sarma}}}, \bibinfo {author} {\bibfnamefont {S.}~\bibnamefont {Adam}},
  \bibinfo {author} {\bibfnamefont {E.~H.}\ \bibnamefont {Hwang}}, \ and\
  \bibinfo {author} {\bibfnamefont {E.}~\bibnamefont {Rossi}},\ }\href@noop {}
  {\bibfield  {journal} {\bibinfo  {journal} {Rev. Mod. Phys.}\ }\textbf
  {\bibinfo {volume} {83}},\ \bibinfo {pages} {407} (\bibinfo {year}
  {2011})}\BibitemShut {NoStop}%
\bibitem [{\citenamefont {Schwierz}(2010)}]{Schwierz}%
  \BibitemOpen
  \bibfield  {author} {\bibinfo {author} {\bibfnamefont {F.}~\bibnamefont
  {Schwierz}},\ }\href@noop {} {\bibfield  {journal} {\bibinfo  {journal}
  {Nature Nanotechnology}\ }\textbf {\bibinfo {volume} {5}},\ \bibinfo {pages}
  {487} (\bibinfo {year} {2010})}\BibitemShut {NoStop}%
\bibitem [{\citenamefont {Morris}\ and\ \citenamefont
  {Iniewski}(2013)}]{morris}%
  \BibitemOpen
  \bibinfo {editor} {\bibfnamefont {J.~E.}\ \bibnamefont {Morris}}\ and\
  \bibinfo {editor} {\bibfnamefont {K.}~\bibnamefont {Iniewski}},\ eds.,\
  \href@noop {} {\emph {\bibinfo {title} {Graphene, Carbon Nanotubes, and
  Nanostructures: Techniques and Applications}}}\ (\bibinfo  {publisher} {CRC
  Press},\ \bibinfo {address} {Florida},\ \bibinfo {year} {2013})\BibitemShut
  {NoStop}%
\bibitem [{\citenamefont {Berger}\ \emph {et~al.}(2006)\citenamefont {Berger},
  \citenamefont {Song}, \citenamefont {Li}, \citenamefont {Wu}, \citenamefont
  {Brown}, \citenamefont {Naud}, \citenamefont {Mayou}, \citenamefont {Li},
  \citenamefont {Hass}, \citenamefont {Marchenkov}, \citenamefont {Conrad},
  \citenamefont {First},\ and\ \citenamefont {de~Heer}}]{berger}%
  \BibitemOpen
  \bibfield  {author} {\bibinfo {author} {\bibfnamefont {C.}~\bibnamefont
  {Berger}}, \bibinfo {author} {\bibfnamefont {Z.}~\bibnamefont {Song}},
  \bibinfo {author} {\bibfnamefont {X.}~\bibnamefont {Li}}, \bibinfo {author}
  {\bibfnamefont {X.}~\bibnamefont {Wu}}, \bibinfo {author} {\bibfnamefont
  {N.}~\bibnamefont {Brown}}, \bibinfo {author} {\bibfnamefont
  {C.}~\bibnamefont {Naud}}, \bibinfo {author} {\bibfnamefont {D.}~\bibnamefont
  {Mayou}}, \bibinfo {author} {\bibfnamefont {T.}~\bibnamefont {Li}}, \bibinfo
  {author} {\bibfnamefont {J.}~\bibnamefont {Hass}}, \bibinfo {author}
  {\bibfnamefont {A.~N.}\ \bibnamefont {Marchenkov}}, \bibinfo {author}
  {\bibfnamefont {E.~H.}\ \bibnamefont {Conrad}}, \bibinfo {author}
  {\bibfnamefont {P.~N.}\ \bibnamefont {First}}, \ and\ \bibinfo {author}
  {\bibfnamefont {W.~A.}\ \bibnamefont {de~Heer}},\ }\href@noop {} {\bibfield
  {journal} {\bibinfo  {journal} {Science}\ }\textbf {\bibinfo {volume}
  {312}},\ \bibinfo {pages} {1191} (\bibinfo {year} {2006})}\BibitemShut
  {NoStop}%
\bibitem [{\citenamefont {Chen}\ \emph {et~al.}(2007)\citenamefont {Chen},
  \citenamefont {Lin}, \citenamefont {Rooks},\ and\ \citenamefont
  {Avouris}}]{chen}%
  \BibitemOpen
  \bibfield  {author} {\bibinfo {author} {\bibfnamefont {Z.}~\bibnamefont
  {Chen}}, \bibinfo {author} {\bibfnamefont {Y.-M.}\ \bibnamefont {Lin}},
  \bibinfo {author} {\bibfnamefont {M.~J.}\ \bibnamefont {Rooks}}, \ and\
  \bibinfo {author} {\bibfnamefont {P.}~\bibnamefont {Avouris}},\ }\href@noop
  {} {\bibfield  {journal} {\bibinfo  {journal} {Physica E}\ }\textbf {\bibinfo
  {volume} {40}},\ \bibinfo {pages} {228} (\bibinfo {year} {2007})}\BibitemShut
  {NoStop}%
\bibitem [{\citenamefont {Han}\ \emph {et~al.}(2007)\citenamefont {Han},
  \citenamefont {{\"Ozyilmaz}}, \citenamefont {Zhang},\ and\ \citenamefont
  {Kim}}]{han}%
  \BibitemOpen
  \bibfield  {author} {\bibinfo {author} {\bibfnamefont {M.~Y.}\ \bibnamefont
  {Han}}, \bibinfo {author} {\bibfnamefont {B.}~\bibnamefont {{\"Ozyilmaz}}},
  \bibinfo {author} {\bibfnamefont {Y.}~\bibnamefont {Zhang}}, \ and\ \bibinfo
  {author} {\bibfnamefont {P.}~\bibnamefont {Kim}},\ }\href@noop {} {\bibfield
  {journal} {\bibinfo  {journal} {Phys. Rev. Lett.}\ }\textbf {\bibinfo
  {volume} {98}},\ \bibinfo {pages} {206805} (\bibinfo {year}
  {2007})}\BibitemShut {NoStop}%
\bibitem [{\citenamefont {Tapaszt\'o}\ \emph {et~al.}(2008)\citenamefont
  {Tapaszt\'o}, \citenamefont {Dobrik}, \citenamefont {Lambin},\ and\
  \citenamefont {Bir\'o}}]{tapaszto}%
  \BibitemOpen
  \bibfield  {author} {\bibinfo {author} {\bibfnamefont {L.}~\bibnamefont
  {Tapaszt\'o}}, \bibinfo {author} {\bibfnamefont {G.}~\bibnamefont {Dobrik}},
  \bibinfo {author} {\bibfnamefont {P.}~\bibnamefont {Lambin}}, \ and\ \bibinfo
  {author} {\bibfnamefont {L.~P.}\ \bibnamefont {Bir\'o}},\ }\href@noop {}
  {\bibfield  {journal} {\bibinfo  {journal} {Nature Nanotechnology}\ }\textbf
  {\bibinfo {volume} {3}},\ \bibinfo {pages} {397} (\bibinfo {year}
  {2008})}\BibitemShut {NoStop}%
\bibitem [{\citenamefont {Datta}\ \emph {et~al.}(2008)\citenamefont {Datta},
  \citenamefont {Strachan}, \citenamefont {Khamis},\ and\ \citenamefont
  {Johnson}}]{datta1}%
  \BibitemOpen
  \bibfield  {author} {\bibinfo {author} {\bibfnamefont {S.~S.}\ \bibnamefont
  {Datta}}, \bibinfo {author} {\bibfnamefont {D.~R.}\ \bibnamefont {Strachan}},
  \bibinfo {author} {\bibfnamefont {S.~M.}\ \bibnamefont {Khamis}}, \ and\
  \bibinfo {author} {\bibfnamefont {A.~T.~C.}\ \bibnamefont {Johnson}},\
  }\href@noop {} {\bibfield  {journal} {\bibinfo  {journal} {Nano Lett.}\
  }\textbf {\bibinfo {volume} {8}},\ \bibinfo {pages} {1912} (\bibinfo {year}
  {2008})}\BibitemShut {NoStop}%
\bibitem [{\citenamefont {Li}\ \emph {et~al.}(2008)\citenamefont {Li},
  \citenamefont {Wang}, \citenamefont {Zhang}, \citenamefont {Lee},\ and\
  \citenamefont {Dai}}]{li}%
  \BibitemOpen
  \bibfield  {author} {\bibinfo {author} {\bibfnamefont {X.}~\bibnamefont
  {Li}}, \bibinfo {author} {\bibfnamefont {X.}~\bibnamefont {Wang}}, \bibinfo
  {author} {\bibfnamefont {L.}~\bibnamefont {Zhang}}, \bibinfo {author}
  {\bibfnamefont {S.}~\bibnamefont {Lee}}, \ and\ \bibinfo {author}
  {\bibfnamefont {H.}~\bibnamefont {Dai}},\ }\href@noop {} {\bibfield
  {journal} {\bibinfo  {journal} {Science}\ }\textbf {\bibinfo {volume}
  {319}},\ \bibinfo {pages} {1229} (\bibinfo {year} {2008})}\BibitemShut
  {NoStop}%
\bibitem [{\citenamefont {Fujita}\ \emph {et~al.}(1996)\citenamefont {Fujita},
  \citenamefont {Wakabayashi}, \citenamefont {Nakada},\ and\ \citenamefont
  {Kusakabe}}]{fujita}%
  \BibitemOpen
  \bibfield  {author} {\bibinfo {author} {\bibfnamefont {M.}~\bibnamefont
  {Fujita}}, \bibinfo {author} {\bibfnamefont {K.}~\bibnamefont {Wakabayashi}},
  \bibinfo {author} {\bibfnamefont {K.}~\bibnamefont {Nakada}}, \ and\ \bibinfo
  {author} {\bibfnamefont {K.}~\bibnamefont {Kusakabe}},\ }\href@noop {}
  {\bibfield  {journal} {\bibinfo  {journal} {J. Phys. Soc. Jpn.}\ }\textbf
  {\bibinfo {volume} {65}},\ \bibinfo {pages} {1920} (\bibinfo {year}
  {1996})}\BibitemShut {NoStop}%
\bibitem [{\citenamefont {Nakada}\ \emph {et~al.}(1996)\citenamefont {Nakada},
  \citenamefont {Fujita}, \citenamefont {Dresselhaus},\ and\ \citenamefont
  {Dresselhaus}}]{nakada}%
  \BibitemOpen
  \bibfield  {author} {\bibinfo {author} {\bibfnamefont {K.}~\bibnamefont
  {Nakada}}, \bibinfo {author} {\bibfnamefont {M.}~\bibnamefont {Fujita}},
  \bibinfo {author} {\bibfnamefont {G.}~\bibnamefont {Dresselhaus}}, \ and\
  \bibinfo {author} {\bibfnamefont {M.~S.}\ \bibnamefont {Dresselhaus}},\
  }\href@noop {} {\bibfield  {journal} {\bibinfo  {journal} {Phys. Rev. B}\
  }\textbf {\bibinfo {volume} {54}},\ \bibinfo {pages} {17954} (\bibinfo {year}
  {1996})}\BibitemShut {NoStop}%
\bibitem [{\citenamefont {Ezawa}(2006)}]{ezawa}%
  \BibitemOpen
  \bibfield  {author} {\bibinfo {author} {\bibfnamefont {M.}~\bibnamefont
  {Ezawa}},\ }\href@noop {} {\bibfield  {journal} {\bibinfo  {journal} {Phys.
  Rev. B}\ }\textbf {\bibinfo {volume} {73}},\ \bibinfo {pages} {045432}
  (\bibinfo {year} {2006})}\BibitemShut {NoStop}%
\bibitem [{\citenamefont {Wakabayashi}\ and\ \citenamefont
  {Dutta}(2012)}]{Wakabayashi}%
  \BibitemOpen
  \bibfield  {author} {\bibinfo {author} {\bibfnamefont {K.}~\bibnamefont
  {Wakabayashi}}\ and\ \bibinfo {author} {\bibfnamefont {S.}~\bibnamefont
  {Dutta}},\ }\href@noop {} {\bibfield  {journal} {\bibinfo  {journal} {Solid
  State Commun.}\ }\textbf {\bibinfo {volume} {152}},\ \bibinfo {pages} {1420}
  (\bibinfo {year} {2012})}\BibitemShut {NoStop}%
\bibitem [{\citenamefont {Brey}\ and\ \citenamefont {Fertig}(2006)}]{brey}%
  \BibitemOpen
  \bibfield  {author} {\bibinfo {author} {\bibfnamefont {L.}~\bibnamefont
  {Brey}}\ and\ \bibinfo {author} {\bibfnamefont {H.~A.}\ \bibnamefont
  {Fertig}},\ }\href@noop {} {\bibfield  {journal} {\bibinfo  {journal} {Phys.
  Rev. B}\ }\textbf {\bibinfo {volume} {73}},\ \bibinfo {pages} {235411}
  (\bibinfo {year} {2006})}\BibitemShut {NoStop}%
\bibitem [{\citenamefont {Marinopoulos}\ \emph {et~al.}(2003)\citenamefont
  {Marinopoulos}, \citenamefont {Reining}, \citenamefont {Rubio},\ and\
  \citenamefont {Vast}}]{marinopoulos03}%
  \BibitemOpen
  \bibfield  {author} {\bibinfo {author} {\bibfnamefont {A.~G.}\ \bibnamefont
  {Marinopoulos}}, \bibinfo {author} {\bibfnamefont {L.}~\bibnamefont
  {Reining}}, \bibinfo {author} {\bibfnamefont {A.}~\bibnamefont {Rubio}}, \
  and\ \bibinfo {author} {\bibfnamefont {N.}~\bibnamefont {Vast}},\ }\href@noop
  {} {\bibfield  {journal} {\bibinfo  {journal} {Phys. Rev. Lett.}\ }\textbf
  {\bibinfo {volume} {91}},\ \bibinfo {pages} {046402} (\bibinfo {year}
  {2003})}\BibitemShut {NoStop}%
\bibitem [{\citenamefont {Marinopoulos}\ \emph {et~al.}(2004)\citenamefont
  {Marinopoulos}, \citenamefont {Wirtz}, \citenamefont {Marini}, \citenamefont
  {Olevano}, \citenamefont {Rubio},\ and\ \citenamefont
  {Reining}}]{marinopoulos04}%
  \BibitemOpen
  \bibfield  {author} {\bibinfo {author} {\bibfnamefont {A.~G.}\ \bibnamefont
  {Marinopoulos}}, \bibinfo {author} {\bibfnamefont {L.}~\bibnamefont {Wirtz}},
  \bibinfo {author} {\bibfnamefont {A.}~\bibnamefont {Marini}}, \bibinfo
  {author} {\bibfnamefont {V.}~\bibnamefont {Olevano}}, \bibinfo {author}
  {\bibfnamefont {A.}~\bibnamefont {Rubio}}, \ and\ \bibinfo {author}
  {\bibfnamefont {L.}~\bibnamefont {Reining}},\ }\href@noop {} {\bibfield
  {journal} {\bibinfo  {journal} {Appl. Phys. A}\ }\textbf {\bibinfo {volume}
  {78}},\ \bibinfo {pages} {1157} (\bibinfo {year} {2004})}\BibitemShut
  {NoStop}%
\bibitem [{\citenamefont {Tachikawa}\ \emph {et~al.}(2009)\citenamefont
  {Tachikawa}, \citenamefont {Nagoya},\ and\ \citenamefont
  {Kawabata}}]{tachikawa09}%
  \BibitemOpen
  \bibfield  {author} {\bibinfo {author} {\bibfnamefont {H.}~\bibnamefont
  {Tachikawa}}, \bibinfo {author} {\bibfnamefont {Y.}~\bibnamefont {Nagoya}}, \
  and\ \bibinfo {author} {\bibfnamefont {H.}~\bibnamefont {Kawabata}},\
  }\href@noop {} {\bibfield  {journal} {\bibinfo  {journal} {J. Chem. Theory
  Comput.}\ }\textbf {\bibinfo {volume} {5}},\ \bibinfo {pages} {2101–2107}
  (\bibinfo {year} {2009})}\BibitemShut {NoStop}%
\bibitem [{\citenamefont {Habenicht}\ and\ \citenamefont
  {Prezhdo}(2009)}]{habenicht09}%
  \BibitemOpen
  \bibfield  {author} {\bibinfo {author} {\bibfnamefont {B.~F.}\ \bibnamefont
  {Habenicht}}\ and\ \bibinfo {author} {\bibfnamefont {O.~V.}\ \bibnamefont
  {Prezhdo}},\ }\href@noop {} {\bibfield  {journal} {\bibinfo  {journal} {J.
  Phys. Chem. C}\ }\textbf {\bibinfo {volume} {113}},\ \bibinfo {pages} {14067}
  (\bibinfo {year} {2009})}\BibitemShut {NoStop}%
\bibitem [{\citenamefont {Spataru}\ \emph {et~al.}(2004)\citenamefont
  {Spataru}, \citenamefont {Ismail-Beigi}, \citenamefont {Benedict},\ and\
  \citenamefont {Louie}}]{spataru04}%
  \BibitemOpen
  \bibfield  {author} {\bibinfo {author} {\bibfnamefont {C.~D.}\ \bibnamefont
  {Spataru}}, \bibinfo {author} {\bibfnamefont {S.}~\bibnamefont
  {Ismail-Beigi}}, \bibinfo {author} {\bibfnamefont {L.~X.}\ \bibnamefont
  {Benedict}}, \ and\ \bibinfo {author} {\bibfnamefont {S.~G.}\ \bibnamefont
  {Louie}},\ }\href@noop {} {\bibfield  {journal} {\bibinfo  {journal} {Phys.
  Rev. Lett.}\ }\textbf {\bibinfo {volume} {92}},\ \bibinfo {pages} {077402}
  (\bibinfo {year} {2004})}\BibitemShut {NoStop}%
\bibitem [{\citenamefont {Capaz}\ \emph {et~al.}(2005)\citenamefont {Capaz},
  \citenamefont {Spataru}, \citenamefont {Tangney}, \citenamefont {Cohen},\
  and\ \citenamefont {Louie}}]{capaz05}%
  \BibitemOpen
  \bibfield  {author} {\bibinfo {author} {\bibfnamefont {R.~B.}\ \bibnamefont
  {Capaz}}, \bibinfo {author} {\bibfnamefont {C.~D.}\ \bibnamefont {Spataru}},
  \bibinfo {author} {\bibfnamefont {P.}~\bibnamefont {Tangney}}, \bibinfo
  {author} {\bibfnamefont {M.~L.}\ \bibnamefont {Cohen}}, \ and\ \bibinfo
  {author} {\bibfnamefont {S.~G.}\ \bibnamefont {Louie}},\ }\href@noop {}
  {\bibfield  {journal} {\bibinfo  {journal} {Phys. Rev. Lett.}\ }\textbf
  {\bibinfo {volume} {94}},\ \bibinfo {pages} {036801} (\bibinfo {year}
  {2005})}\BibitemShut {NoStop}%
\bibitem [{\citenamefont {Yang}\ \emph {et~al.}(2007)\citenamefont {Yang},
  \citenamefont {Park}, \citenamefont {Son}, \citenamefont {Cohen},\ and\
  \citenamefont {Louie}}]{yang07}%
  \BibitemOpen
  \bibfield  {author} {\bibinfo {author} {\bibfnamefont {L.}~\bibnamefont
  {Yang}}, \bibinfo {author} {\bibfnamefont {C.-H.}\ \bibnamefont {Park}},
  \bibinfo {author} {\bibfnamefont {Y.-W.}\ \bibnamefont {Son}}, \bibinfo
  {author} {\bibfnamefont {M.~L.}\ \bibnamefont {Cohen}}, \ and\ \bibinfo
  {author} {\bibfnamefont {S.~G.}\ \bibnamefont {Louie}},\ }\href@noop {}
  {\bibfield  {journal} {\bibinfo  {journal} {Phys. Rev. Lett.}\ }\textbf
  {\bibinfo {volume} {99}},\ \bibinfo {pages} {186801} (\bibinfo {year}
  {2007})}\BibitemShut {NoStop}%
\bibitem [{\citenamefont {Zhao}\ and\ \citenamefont {Mazumdar}(2004)}]{Zhao04}%
  \BibitemOpen
  \bibfield  {author} {\bibinfo {author} {\bibfnamefont {H.}~\bibnamefont
  {Zhao}}\ and\ \bibinfo {author} {\bibfnamefont {S.}~\bibnamefont
  {Mazumdar}},\ }\href@noop {} {\bibfield  {journal} {\bibinfo  {journal}
  {Phys. Rev. Lett.}\ }\textbf {\bibinfo {volume} {93}},\ \bibinfo {pages}
  {157402} (\bibinfo {year} {2004})}\BibitemShut {NoStop}%
\bibitem [{\citenamefont {Wang}\ \emph {et~al.}(2006)\citenamefont {Wang},
  \citenamefont {Zhao},\ and\ \citenamefont {Mazumdar}}]{Wang06}%
  \BibitemOpen
  \bibfield  {author} {\bibinfo {author} {\bibfnamefont {Z.}~\bibnamefont
  {Wang}}, \bibinfo {author} {\bibfnamefont {H.}~\bibnamefont {Zhao}}, \ and\
  \bibinfo {author} {\bibfnamefont {S.}~\bibnamefont {Mazumdar}},\ }\href@noop
  {} {\bibfield  {journal} {\bibinfo  {journal} {Phys. Rev. B}\ }\textbf
  {\bibinfo {volume} {74}},\ \bibinfo {pages} {195406} (\bibinfo {year}
  {2006})}\BibitemShut {NoStop}%
\bibitem [{\citenamefont {Wang}\ \emph {et~al.}(2007)\citenamefont {Wang},
  \citenamefont {Zhao},\ and\ \citenamefont {Mazumdar}}]{Wang07}%
  \BibitemOpen
  \bibfield  {author} {\bibinfo {author} {\bibfnamefont {Z.}~\bibnamefont
  {Wang}}, \bibinfo {author} {\bibfnamefont {H.}~\bibnamefont {Zhao}}, \ and\
  \bibinfo {author} {\bibfnamefont {S.}~\bibnamefont {Mazumdar}},\ }\href@noop
  {} {\bibfield  {journal} {\bibinfo  {journal} {Phys. Rev. B}\ }\textbf
  {\bibinfo {volume} {76}},\ \bibinfo {pages} {115431} (\bibinfo {year}
  {2007})}\BibitemShut {NoStop}%
\bibitem [{\citenamefont {Hudson}\ \emph {et~al.}(1982)\citenamefont {Hudson},
  \citenamefont {Kohler},\ and\ \citenamefont
  {Schulten}}]{hudson-kohler-schulten}%
  \BibitemOpen
  \bibfield  {author} {\bibinfo {author} {\bibfnamefont {B.~S.}\ \bibnamefont
  {Hudson}}, \bibinfo {author} {\bibfnamefont {B.~E.}\ \bibnamefont {Kohler}},
  \ and\ \bibinfo {author} {\bibfnamefont {K.}~\bibnamefont {Schulten}},\
  }\href@noop {} {\emph {\bibinfo {title} {Excited States}}},\ edited by\
  \bibinfo {editor} {\bibfnamefont {E.~C.}\ \bibnamefont {Lim}},\ Vol.~\bibinfo
  {volume} {6}\ (\bibinfo  {publisher} {Academic Press},\ \bibinfo {address}
  {New York},\ \bibinfo {year} {1982})\ p.~\bibinfo {pages} {1}\BibitemShut
  {NoStop}%
\bibitem [{\citenamefont {Aryanpour}\ \emph {et~al.}(2014)\citenamefont
  {Aryanpour}, \citenamefont {Roberts}, \citenamefont {Sandhu}, \citenamefont
  {Rathore}, \citenamefont {Shukla},\ and\ \citenamefont
  {Mazumdar}}]{Sandhu-Mazumdar}%
  \BibitemOpen
  \bibfield  {author} {\bibinfo {author} {\bibfnamefont {K.}~\bibnamefont
  {Aryanpour}}, \bibinfo {author} {\bibfnamefont {A.}~\bibnamefont {Roberts}},
  \bibinfo {author} {\bibfnamefont {A.}~\bibnamefont {Sandhu}}, \bibinfo
  {author} {\bibfnamefont {R.}~\bibnamefont {Rathore}}, \bibinfo {author}
  {\bibfnamefont {A.}~\bibnamefont {Shukla}}, \ and\ \bibinfo {author}
  {\bibfnamefont {S.}~\bibnamefont {Mazumdar}},\ }\href@noop {} {\bibfield
  {journal} {\bibinfo  {journal} {J. Phys. Chem. C}\ }\textbf {\bibinfo
  {volume} {118}},\ \bibinfo {pages} {3331} (\bibinfo {year}
  {2014})}\BibitemShut {NoStop}%
\bibitem [{\citenamefont {Pariser}\ and\ \citenamefont {Parr}(1953)}]{pppm}%
  \BibitemOpen
  \bibfield  {author} {\bibinfo {author} {\bibfnamefont {R.}~\bibnamefont
  {Pariser}}\ and\ \bibinfo {author} {\bibfnamefont {R.~G.}\ \bibnamefont
  {Parr}},\ }\href@noop {} {\bibfield  {journal} {\bibinfo  {journal} {J. Chem.
  Phys.}\ }\textbf {\bibinfo {volume} {21}},\ \bibinfo {pages} {466} (\bibinfo
  {year} {1953})}\BibitemShut {NoStop}%
\bibitem [{\citenamefont {Pople}(1953)}]{pople}%
  \BibitemOpen
  \bibfield  {author} {\bibinfo {author} {\bibfnamefont {J.~A.}\ \bibnamefont
  {Pople}},\ }\href@noop {} {\bibfield  {journal} {\bibinfo  {journal} {Trans.
  Faraday Soc.}\ }\textbf {\bibinfo {volume} {49}},\ \bibinfo {pages} {1375}
  (\bibinfo {year} {1953})}\BibitemShut {NoStop}%
\bibitem [{\citenamefont {Tavan}\ and\ \citenamefont
  {Schulten}(1979{\natexlab{a}})}]{schulten79-1}%
  \BibitemOpen
  \bibfield  {author} {\bibinfo {author} {\bibfnamefont {P.}~\bibnamefont
  {Tavan}}\ and\ \bibinfo {author} {\bibfnamefont {K.}~\bibnamefont
  {Schulten}},\ }\href@noop {} {\bibfield  {journal} {\bibinfo  {journal} {J.
  Chem. Phys.}\ }\textbf {\bibinfo {volume} {70}},\ \bibinfo {pages} {5407}
  (\bibinfo {year} {1979}{\natexlab{a}})}\BibitemShut {NoStop}%
\bibitem [{\citenamefont {Tavan}\ and\ \citenamefont
  {Schulten}(1979{\natexlab{b}})}]{schulten79-2}%
  \BibitemOpen
  \bibfield  {author} {\bibinfo {author} {\bibfnamefont {P.}~\bibnamefont
  {Tavan}}\ and\ \bibinfo {author} {\bibfnamefont {K.}~\bibnamefont
  {Schulten}},\ }\href@noop {} {\bibfield  {journal} {\bibinfo  {journal} {J.
  Chem. Phys.}\ }\textbf {\bibinfo {volume} {70}},\ \bibinfo {pages} {5414}
  (\bibinfo {year} {1979}{\natexlab{b}})}\BibitemShut {NoStop}%
\bibitem [{\citenamefont {Tavan}\ and\ \citenamefont
  {Schulten}(1986)}]{schulten86}%
  \BibitemOpen
  \bibfield  {author} {\bibinfo {author} {\bibfnamefont {P.}~\bibnamefont
  {Tavan}}\ and\ \bibinfo {author} {\bibfnamefont {K.}~\bibnamefont
  {Schulten}},\ }\href@noop {} {\bibfield  {journal} {\bibinfo  {journal} {J.
  Chem. Phys.}\ }\textbf {\bibinfo {volume} {85}},\ \bibinfo {pages} {6602}
  (\bibinfo {year} {1986})}\BibitemShut {NoStop}%
\bibitem [{\citenamefont {Tavan}\ and\ \citenamefont
  {Schulten}(1987)}]{schulten87}%
  \BibitemOpen
  \bibfield  {author} {\bibinfo {author} {\bibfnamefont {P.}~\bibnamefont
  {Tavan}}\ and\ \bibinfo {author} {\bibfnamefont {K.}~\bibnamefont
  {Schulten}},\ }\href@noop {} {\bibfield  {journal} {\bibinfo  {journal}
  {Phys. Rev. B}\ }\textbf {\bibinfo {volume} {36}},\ \bibinfo {pages} {4337}
  (\bibinfo {year} {1987})}\BibitemShut {NoStop}%
\bibitem [{\citenamefont {Soos}\ \emph {et~al.}(1993)\citenamefont {Soos},
  \citenamefont {Ramasesha},\ and\ \citenamefont {Galvao}}]{soos93}%
  \BibitemOpen
  \bibfield  {author} {\bibinfo {author} {\bibfnamefont {Z.~G.}\ \bibnamefont
  {Soos}}, \bibinfo {author} {\bibfnamefont {S.}~\bibnamefont {Ramasesha}}, \
  and\ \bibinfo {author} {\bibfnamefont {D.~S.}\ \bibnamefont {Galvao}},\
  }\href@noop {} {\bibfield  {journal} {\bibinfo  {journal} {Phys. Rev. Lett.}\
  }\textbf {\bibinfo {volume} {71}},\ \bibinfo {pages} {1609} (\bibinfo {year}
  {1993})}\BibitemShut {NoStop}%
\bibitem [{\citenamefont {Soos}\ and\ \citenamefont {Ramasesha}(1984)}]{soos}%
  \BibitemOpen
  \bibfield  {author} {\bibinfo {author} {\bibfnamefont {Z.~G.}\ \bibnamefont
  {Soos}}\ and\ \bibinfo {author} {\bibfnamefont {S.}~\bibnamefont
  {Ramasesha}},\ }\href@noop {} {\bibfield  {journal} {\bibinfo  {journal}
  {Phys. Rev. B}\ }\textbf {\bibinfo {volume} {29}},\ \bibinfo {pages} {5410}
  (\bibinfo {year} {1984})}\BibitemShut {NoStop}%
\bibitem [{\citenamefont {Baeriswyl}\ \emph {et~al.}(1992)\citenamefont
  {Baeriswyl}, \citenamefont {Campbell},\ and\ \citenamefont
  {Mazumdar}}]{baeriswyl}%
  \BibitemOpen
  \bibfield  {author} {\bibinfo {author} {\bibfnamefont {D.}~\bibnamefont
  {Baeriswyl}}, \bibinfo {author} {\bibfnamefont {D.~K.}\ \bibnamefont
  {Campbell}}, \ and\ \bibinfo {author} {\bibfnamefont {S.}~\bibnamefont
  {Mazumdar}},\ }\href@noop {} {\emph {\bibinfo {title} {Conjugated Conducting
  Polymers}}},\ edited by\ \bibinfo {editor} {\bibfnamefont {H.}~\bibnamefont
  {Kiess}},\ \bibinfo {series} {Springer Series in Solid-State Sciences}, Vol.\
  \bibinfo {volume} {102}\ (\bibinfo  {publisher} {Springer},\ \bibinfo
  {address} {Berlin},\ \bibinfo {year} {1992})\BibitemShut {NoStop}%
\bibitem [{\citenamefont {Raghu}\ \emph
  {et~al.}(2002{\natexlab{a}})\citenamefont {Raghu}, \citenamefont {{Anusooya
  Pati}},\ and\ \citenamefont {Ramasesha}}]{raghu02-1}%
  \BibitemOpen
  \bibfield  {author} {\bibinfo {author} {\bibfnamefont {C.}~\bibnamefont
  {Raghu}}, \bibinfo {author} {\bibfnamefont {Y.}~\bibnamefont {{Anusooya
  Pati}}}, \ and\ \bibinfo {author} {\bibfnamefont {S.}~\bibnamefont
  {Ramasesha}},\ }\href@noop {} {\bibfield  {journal} {\bibinfo  {journal}
  {Phys. Rev. B}\ }\textbf {\bibinfo {volume} {65}},\ \bibinfo {pages} {155204}
  (\bibinfo {year} {2002}{\natexlab{a}})}\BibitemShut {NoStop}%
\bibitem [{\citenamefont {Kumar}\ and\ \citenamefont
  {Ramasesha}(2010)}]{mano10}%
  \BibitemOpen
  \bibfield  {author} {\bibinfo {author} {\bibfnamefont {M.}~\bibnamefont
  {Kumar}}\ and\ \bibinfo {author} {\bibfnamefont {S.}~\bibnamefont
  {Ramasesha}},\ }\href@noop {} {\bibfield  {journal} {\bibinfo  {journal}
  {Phys. Rev. B}\ }\textbf {\bibinfo {volume} {81}},\ \bibinfo {pages} {035115}
  (\bibinfo {year} {2010})}\BibitemShut {NoStop}%
\bibitem [{\citenamefont {Thomas}\ \emph {et~al.}(2013)\citenamefont {Thomas},
  \citenamefont {Pati},\ and\ \citenamefont {Ramasesha}}]{simil13}%
  \BibitemOpen
  \bibfield  {author} {\bibinfo {author} {\bibfnamefont {S.}~\bibnamefont
  {Thomas}}, \bibinfo {author} {\bibfnamefont {Y.~A.}\ \bibnamefont {Pati}}, \
  and\ \bibinfo {author} {\bibfnamefont {S.}~\bibnamefont {Ramasesha}},\
  }\href@noop {} {\bibfield  {journal} {\bibinfo  {journal} {J. Phys. Chem. A}\
  }\textbf {\bibinfo {volume} {117}},\ \bibinfo {pages} {7804–7809} (\bibinfo
  {year} {2013})}\BibitemShut {NoStop}%
\bibitem [{\citenamefont {Mukhopadhyay}\ and\ \citenamefont
  {Ramasesha}(2009)}]{sukrit09}%
  \BibitemOpen
  \bibfield  {author} {\bibinfo {author} {\bibfnamefont {S.}~\bibnamefont
  {Mukhopadhyay}}\ and\ \bibinfo {author} {\bibfnamefont {S.}~\bibnamefont
  {Ramasesha}},\ }\href@noop {} {\bibfield  {journal} {\bibinfo  {journal} {J.
  Chem. Phys.}\ }\textbf {\bibinfo {volume} {131}},\ \bibinfo {pages} {074111}
  (\bibinfo {year} {2009})}\BibitemShut {NoStop}%
\bibitem [{\citenamefont {Raghu}\ \emph
  {et~al.}(2002{\natexlab{b}})\citenamefont {Raghu}, \citenamefont {{Anusooya
  Pati}},\ and\ \citenamefont {Ramasesha}}]{raghu02-2}%
  \BibitemOpen
  \bibfield  {author} {\bibinfo {author} {\bibfnamefont {C.}~\bibnamefont
  {Raghu}}, \bibinfo {author} {\bibfnamefont {Y.}~\bibnamefont {{Anusooya
  Pati}}}, \ and\ \bibinfo {author} {\bibfnamefont {S.}~\bibnamefont
  {Ramasesha}},\ }\href@noop {} {\bibfield  {journal} {\bibinfo  {journal}
  {Phys. Rev. B}\ }\textbf {\bibinfo {volume} {66}},\ \bibinfo {pages} {035116}
  (\bibinfo {year} {2002}{\natexlab{b}})}\BibitemShut {NoStop}%
\bibitem [{\citenamefont {Sasaki}\ \emph {et~al.}(2006)\citenamefont {Sasaki},
  \citenamefont {Murakami},\ and\ \citenamefont {Saito}}]{sasaki06}%
  \BibitemOpen
  \bibfield  {author} {\bibinfo {author} {\bibfnamefont {K.}~\bibnamefont
  {Sasaki}}, \bibinfo {author} {\bibfnamefont {S.}~\bibnamefont {Murakami}}, \
  and\ \bibinfo {author} {\bibfnamefont {R.}~\bibnamefont {Saito}},\
  }\href@noop {} {\bibfield  {journal} {\bibinfo  {journal} {J. Phys. Soc.
  Jpn}\ }\textbf {\bibinfo {volume} {75}},\ \bibinfo {pages} {074713} (\bibinfo
  {year} {2006})}\BibitemShut {NoStop}%
\bibitem [{\citenamefont {{Castro Neto}}\ \emph {et~al.}(2006)\citenamefont
  {{Castro Neto}}, \citenamefont {Guinea},\ and\ \citenamefont
  {Peres}}]{castroneto06}%
  \BibitemOpen
  \bibfield  {author} {\bibinfo {author} {\bibfnamefont {A.~H.}\ \bibnamefont
  {{Castro Neto}}}, \bibinfo {author} {\bibfnamefont {F.}~\bibnamefont
  {Guinea}}, \ and\ \bibinfo {author} {\bibfnamefont {N.~M.~R.}\ \bibnamefont
  {Peres}},\ }\href@noop {} {\bibfield  {journal} {\bibinfo  {journal} {Phys.
  Rev. B}\ }\textbf {\bibinfo {volume} {73}},\ \bibinfo {pages} {205408}
  (\bibinfo {year} {2006})}\BibitemShut {NoStop}%
\bibitem [{\citenamefont {Abanin}\ \emph {et~al.}(2006)\citenamefont {Abanin},
  \citenamefont {Lee},\ and\ \citenamefont {Levitov}}]{levitov06}%
  \BibitemOpen
  \bibfield  {author} {\bibinfo {author} {\bibfnamefont {D.~A.}\ \bibnamefont
  {Abanin}}, \bibinfo {author} {\bibfnamefont {P.~A.}\ \bibnamefont {Lee}}, \
  and\ \bibinfo {author} {\bibfnamefont {L.~S.}\ \bibnamefont {Levitov}},\
  }\href@noop {} {\bibfield  {journal} {\bibinfo  {journal} {Phys. Rev. Lett.}\
  }\textbf {\bibinfo {volume} {96}},\ \bibinfo {pages} {176803} (\bibinfo
  {year} {2006})}\BibitemShut {NoStop}%
\bibitem [{\citenamefont {Wakabayashi}\ \emph {et~al.}(1999)\citenamefont
  {Wakabayashi}, \citenamefont {Fujita}, \citenamefont {Ajiki},\ and\
  \citenamefont {Sigrist}}]{wakabayashi99}%
  \BibitemOpen
  \bibfield  {author} {\bibinfo {author} {\bibfnamefont {K.}~\bibnamefont
  {Wakabayashi}}, \bibinfo {author} {\bibfnamefont {M.}~\bibnamefont {Fujita}},
  \bibinfo {author} {\bibfnamefont {H.}~\bibnamefont {Ajiki}}, \ and\ \bibinfo
  {author} {\bibfnamefont {M.}~\bibnamefont {Sigrist}},\ }\href@noop {}
  {\bibfield  {journal} {\bibinfo  {journal} {Phys. Rev. B}\ }\textbf {\bibinfo
  {volume} {59}},\ \bibinfo {pages} {8271} (\bibinfo {year}
  {1999})}\BibitemShut {NoStop}%
\bibitem [{\citenamefont {Son}\ \emph {et~al.}(2006{\natexlab{a}})\citenamefont
  {Son}, \citenamefont {Cohen},\ and\ \citenamefont {Louie}}]{son06-2}%
  \BibitemOpen
  \bibfield  {author} {\bibinfo {author} {\bibfnamefont {Y.-W.}\ \bibnamefont
  {Son}}, \bibinfo {author} {\bibfnamefont {M.~L.}\ \bibnamefont {Cohen}}, \
  and\ \bibinfo {author} {\bibfnamefont {S.~G.}\ \bibnamefont {Louie}},\
  }\href@noop {} {\bibfield  {journal} {\bibinfo  {journal} {Nature}\ }\textbf
  {\bibinfo {volume} {444}},\ \bibinfo {pages} {347} (\bibinfo {year}
  {2006}{\natexlab{a}})}\BibitemShut {NoStop}%
\bibitem [{\citenamefont {Jung}\ and\ \citenamefont {MacDonald}(2009)}]{jung}%
  \BibitemOpen
  \bibfield  {author} {\bibinfo {author} {\bibfnamefont {J.}~\bibnamefont
  {Jung}}\ and\ \bibinfo {author} {\bibfnamefont {A.~H.}\ \bibnamefont
  {MacDonald}},\ }\href@noop {} {\bibfield  {journal} {\bibinfo  {journal}
  {Phys. Rev. B}\ }\textbf {\bibinfo {volume} {79}},\ \bibinfo {pages} {235433}
  (\bibinfo {year} {2009})}\BibitemShut {NoStop}%
\bibitem [{\citenamefont {Golor}\ \emph {et~al.}(2013)\citenamefont {Golor},
  \citenamefont {Lang},\ and\ \citenamefont {Wessel}}]{golor}%
  \BibitemOpen
  \bibfield  {author} {\bibinfo {author} {\bibfnamefont {M.}~\bibnamefont
  {Golor}}, \bibinfo {author} {\bibfnamefont {T.~C.}\ \bibnamefont {Lang}}, \
  and\ \bibinfo {author} {\bibfnamefont {S.}~\bibnamefont {Wessel}},\
  }\href@noop {} {\bibfield  {journal} {\bibinfo  {journal} {Phys. Rev. B}\
  }\textbf {\bibinfo {volume} {87}},\ \bibinfo {pages} {155441} (\bibinfo
  {year} {2013})}\BibitemShut {NoStop}%
\bibitem [{\citenamefont {Hikihara}\ \emph {et~al.}(2003)\citenamefont
  {Hikihara}, \citenamefont {Hu}, \citenamefont {Lin},\ and\ \citenamefont
  {Mou}}]{hikihara}%
  \BibitemOpen
  \bibfield  {author} {\bibinfo {author} {\bibfnamefont {T.}~\bibnamefont
  {Hikihara}}, \bibinfo {author} {\bibfnamefont {X.}~\bibnamefont {Hu}},
  \bibinfo {author} {\bibfnamefont {H.-H.}\ \bibnamefont {Lin}}, \ and\
  \bibinfo {author} {\bibfnamefont {C.-Y.}\ \bibnamefont {Mou}},\ }\href@noop
  {} {\bibfield  {journal} {\bibinfo  {journal} {Phys. Rev. B}\ }\textbf
  {\bibinfo {volume} {68}},\ \bibinfo {pages} {035432} (\bibinfo {year}
  {2003})}\BibitemShut {NoStop}%
\bibitem [{\citenamefont {Magda}\ \emph {et~al.}(2014)\citenamefont {Magda},
  \citenamefont {Jin}, \citenamefont {Hagym\'{a}si}, \citenamefont
  {Vancs\'{o}}, \citenamefont {Osv\'{a}th}, \citenamefont {Nemes-Incze},
  \citenamefont {Hwang}, \citenamefont {Bir\'{o}},\ and\ \citenamefont
  {Tapaszt\'{o}}}]{magda}%
  \BibitemOpen
  \bibfield  {author} {\bibinfo {author} {\bibfnamefont {G.~Z.}\ \bibnamefont
  {Magda}}, \bibinfo {author} {\bibfnamefont {X.}~\bibnamefont {Jin}}, \bibinfo
  {author} {\bibfnamefont {I.}~\bibnamefont {Hagym\'{a}si}}, \bibinfo {author}
  {\bibfnamefont {P.}~\bibnamefont {Vancs\'{o}}}, \bibinfo {author}
  {\bibfnamefont {Z.}~\bibnamefont {Osv\'{a}th}}, \bibinfo {author}
  {\bibfnamefont {P.}~\bibnamefont {Nemes-Incze}}, \bibinfo {author}
  {\bibfnamefont {C.}~\bibnamefont {Hwang}}, \bibinfo {author} {\bibfnamefont
  {L.~P.}\ \bibnamefont {Bir\'{o}}}, \ and\ \bibinfo {author} {\bibfnamefont
  {L.}~\bibnamefont {Tapaszt\'{o}}},\ }\href@noop {} {\bibfield  {journal}
  {\bibinfo  {journal} {Nature}\ }\textbf {\bibinfo {volume} {514}},\ \bibinfo
  {pages} {608} (\bibinfo {year} {2014})}\BibitemShut {NoStop}%
\bibitem [{\citenamefont {Okada}\ and\ \citenamefont
  {Oshiyama}(2001)}]{okada01}%
  \BibitemOpen
  \bibfield  {author} {\bibinfo {author} {\bibfnamefont {S.}~\bibnamefont
  {Okada}}\ and\ \bibinfo {author} {\bibfnamefont {A.}~\bibnamefont
  {Oshiyama}},\ }\href@noop {} {\bibfield  {journal} {\bibinfo  {journal}
  {Phys. Rev. Lett.}\ }\textbf {\bibinfo {volume} {87}},\ \bibinfo {pages}
  {146803} (\bibinfo {year} {2001})}\BibitemShut {NoStop}%
\bibitem [{\citenamefont {Dutta}\ and\ \citenamefont {Pati}(2008)}]{dutta08}%
  \BibitemOpen
  \bibfield  {author} {\bibinfo {author} {\bibfnamefont {S.}~\bibnamefont
  {Dutta}}\ and\ \bibinfo {author} {\bibfnamefont {S.~K.}\ \bibnamefont
  {Pati}},\ }\href@noop {} {\bibfield  {journal} {\bibinfo  {journal} {J. Phys.
  Chem. B}\ }\textbf {\bibinfo {volume} {112}},\ \bibinfo {pages} {1333}
  (\bibinfo {year} {2008})}\BibitemShut {NoStop}%
\bibitem [{\citenamefont {Dutta}\ \emph {et~al.}(2008)\citenamefont {Dutta},
  \citenamefont {Lakshmi},\ and\ \citenamefont {Pati}}]{datta2}%
  \BibitemOpen
  \bibfield  {author} {\bibinfo {author} {\bibfnamefont {S.}~\bibnamefont
  {Dutta}}, \bibinfo {author} {\bibfnamefont {S.}~\bibnamefont {Lakshmi}}, \
  and\ \bibinfo {author} {\bibfnamefont {S.~K.}\ \bibnamefont {Pati}},\
  }\href@noop {} {\bibfield  {journal} {\bibinfo  {journal} {Phys. Rev. B}\
  }\textbf {\bibinfo {volume} {77}},\ \bibinfo {pages} {073412} (\bibinfo
  {year} {2008})}\BibitemShut {NoStop}%
\bibitem [{\citenamefont {Son}\ \emph {et~al.}(2006{\natexlab{b}})\citenamefont
  {Son}, \citenamefont {Cohen},\ and\ \citenamefont {Louie}}]{son06-1}%
  \BibitemOpen
  \bibfield  {author} {\bibinfo {author} {\bibfnamefont {Y.-W.}\ \bibnamefont
  {Son}}, \bibinfo {author} {\bibfnamefont {M.~L.}\ \bibnamefont {Cohen}}, \
  and\ \bibinfo {author} {\bibfnamefont {S.~G.}\ \bibnamefont {Louie}},\
  }\href@noop {} {\bibfield  {journal} {\bibinfo  {journal} {Phys. Rev. Lett.}\
  }\textbf {\bibinfo {volume} {97}},\ \bibinfo {pages} {216803} (\bibinfo
  {year} {2006}{\natexlab{b}})}\BibitemShut {NoStop}%
\bibitem [{\citenamefont {Das}(2014)}]{das}%
  \BibitemOpen
  \bibfield  {author} {\bibinfo {author} {\bibfnamefont {M.}~\bibnamefont
  {Das}},\ }\href@noop {} {\bibfield  {journal} {\bibinfo  {journal} {J. Chem.
  Phys.}\ }\textbf {\bibinfo {volume} {140}},\ \bibinfo {pages} {124317}
  (\bibinfo {year} {2014})}\BibitemShut {NoStop}%
\bibitem [{\citenamefont {Ohno}(1964)}]{ohno}%
  \BibitemOpen
  \bibfield  {author} {\bibinfo {author} {\bibfnamefont {K.}~\bibnamefont
  {Ohno}},\ }\href@noop {} {\bibfield  {journal} {\bibinfo  {journal} {Theor.
  Chim. Acta}\ }\textbf {\bibinfo {volume} {2}},\ \bibinfo {pages} {219}
  (\bibinfo {year} {1964})}\BibitemShut {NoStop}%
\bibitem [{\citenamefont {Klopman}(1964)}]{klopman}%
  \BibitemOpen
  \bibfield  {author} {\bibinfo {author} {\bibfnamefont {G.}~\bibnamefont
  {Klopman}},\ }\href@noop {} {\bibfield  {journal} {\bibinfo  {journal} {J.
  Am. Chem. Soc.}\ }\textbf {\bibinfo {volume} {86}},\ \bibinfo {pages} {4550}
  (\bibinfo {year} {1964})}\BibitemShut {NoStop}%
\bibitem [{\citenamefont {Race}\ \emph {et~al.}(2001)\citenamefont {Race},
  \citenamefont {Barford},\ and\ \citenamefont {Bursill}}]{race01}%
  \BibitemOpen
  \bibfield  {author} {\bibinfo {author} {\bibfnamefont {A.}~\bibnamefont
  {Race}}, \bibinfo {author} {\bibfnamefont {W.}~\bibnamefont {Barford}}, \
  and\ \bibinfo {author} {\bibfnamefont {R.~J.}\ \bibnamefont {Bursill}},\
  }\href@noop {} {\bibfield  {journal} {\bibinfo  {journal} {Phys. Rev. B}\
  }\textbf {\bibinfo {volume} {64}},\ \bibinfo {pages} {035208} (\bibinfo
  {year} {2001})}\BibitemShut {NoStop}%
\bibitem [{\citenamefont {Race}\ \emph {et~al.}(2003)\citenamefont {Race},
  \citenamefont {Barford},\ and\ \citenamefont {Bursill}}]{race03}%
  \BibitemOpen
  \bibfield  {author} {\bibinfo {author} {\bibfnamefont {A.}~\bibnamefont
  {Race}}, \bibinfo {author} {\bibfnamefont {W.}~\bibnamefont {Barford}}, \
  and\ \bibinfo {author} {\bibfnamefont {R.~J.}\ \bibnamefont {Bursill}},\
  }\href@noop {} {\bibfield  {journal} {\bibinfo  {journal} {Phys. Rev. B}\
  }\textbf {\bibinfo {volume} {67}},\ \bibinfo {pages} {245202} (\bibinfo
  {year} {2003})}\BibitemShut {NoStop}%
\bibitem [{\citenamefont {Salem}(1966)}]{salem}%
  \BibitemOpen
  \bibfield  {author} {\bibinfo {author} {\bibfnamefont {L.}~\bibnamefont
  {Salem}},\ }\href@noop {} {\emph {\bibinfo {title} {The Molecular Orbital
  Theory of Conjugated Systems}}}\ (\bibinfo  {publisher} {W. A. Benjamin,
  Inc.},\ \bibinfo {address} {Massachusetts},\ \bibinfo {year} {1966})\ p.\
  \bibinfo {pages} {420}\BibitemShut {NoStop}%
\bibitem [{\citenamefont {{Schollw\"{o}ck}}(2005)}]{scholl}%
  \BibitemOpen
  \bibfield  {author} {\bibinfo {author} {\bibfnamefont {U.}~\bibnamefont
  {{Schollw\"{o}ck}}},\ }\href@noop {} {\bibfield  {journal} {\bibinfo
  {journal} {Rev. Mod. Phys.}\ }\textbf {\bibinfo {volume} {77}},\ \bibinfo
  {pages} {259} (\bibinfo {year} {2005})}\BibitemShut {NoStop}%
\bibitem [{\citenamefont {Hallberg}(2006)}]{hall}%
  \BibitemOpen
  \bibfield  {author} {\bibinfo {author} {\bibfnamefont {K.~A.}\ \bibnamefont
  {Hallberg}},\ }\href@noop {} {\bibfield  {journal} {\bibinfo  {journal} {Adv.
  Phys.}\ }\textbf {\bibinfo {volume} {55}},\ \bibinfo {pages} {477} (\bibinfo
  {year} {2006})}\BibitemShut {NoStop}%
\bibitem [{\citenamefont {Sahoo}\ \emph {et~al.}(2012)\citenamefont {Sahoo},
  \citenamefont {Goli}, \citenamefont {Ramasesha},\ and\ \citenamefont
  {Sen}}]{durga}%
  \BibitemOpen
  \bibfield  {author} {\bibinfo {author} {\bibfnamefont {S.}~\bibnamefont
  {Sahoo}}, \bibinfo {author} {\bibfnamefont {V.~M. L. D.~P.}\ \bibnamefont
  {Goli}}, \bibinfo {author} {\bibfnamefont {S.}~\bibnamefont {Ramasesha}}, \
  and\ \bibinfo {author} {\bibfnamefont {D.}~\bibnamefont {Sen}},\ }\href@noop
  {} {\bibfield  {journal} {\bibinfo  {journal} {J. Phys.:Condens. Matter}\
  }\textbf {\bibinfo {volume} {24}},\ \bibinfo {pages} {115601} (\bibinfo
  {year} {2012})}\BibitemShut {NoStop}%
\bibitem [{sup()}]{supple}%
  \BibitemOpen
  \href@noop {} {}\bibinfo {note} {See Supplemental Material for the method of
  constructing the AGNRs in the infinite DMRG procedure as well as the block
  building scheme of finite sweeps for 6-AGNR and 5-AGNR.}\BibitemShut {Stop}%
\bibitem [{\citenamefont {White}(1992)}]{white}%
  \BibitemOpen
  \bibfield  {author} {\bibinfo {author} {\bibfnamefont {S.~R.}\ \bibnamefont
  {White}},\ }\href@noop {} {\bibfield  {journal} {\bibinfo  {journal} {Phys.
  Rev. Lett.}\ }\textbf {\bibinfo {volume} {69}},\ \bibinfo {pages} {2863}
  (\bibinfo {year} {1992})}\BibitemShut {NoStop}%
\bibitem [{\citenamefont {White}(1993)}]{white-prb}%
  \BibitemOpen
  \bibfield  {author} {\bibinfo {author} {\bibfnamefont {S.~R.}\ \bibnamefont
  {White}},\ }\href@noop {} {\bibfield  {journal} {\bibinfo  {journal} {Phys.
  Rev. B}\ }\textbf {\bibinfo {volume} {48}},\ \bibinfo {pages} {10345}
  (\bibinfo {year} {1993})}\BibitemShut {NoStop}%
\bibitem [{\citenamefont {Ramasesha}\ \emph {et~al.}(2000)\citenamefont
  {Ramasesha}, \citenamefont {Pati}, \citenamefont {Shuai},\ and\ \citenamefont
  {{Br\'{e}das}}}]{ramasesha}%
  \BibitemOpen
  \bibfield  {author} {\bibinfo {author} {\bibfnamefont {S.}~\bibnamefont
  {Ramasesha}}, \bibinfo {author} {\bibfnamefont {S.~K.}\ \bibnamefont {Pati}},
  \bibinfo {author} {\bibfnamefont {Z.}~\bibnamefont {Shuai}}, \ and\ \bibinfo
  {author} {\bibfnamefont {J.~L.}\ \bibnamefont {{Br\'{e}das}}},\ }\href@noop
  {} {\bibfield  {journal} {\bibinfo  {journal} {Adv. Quantum Chem.}\ }\textbf
  {\bibinfo {volume} {38}},\ \bibinfo {pages} {121} (\bibinfo {year}
  {2000})}\BibitemShut {NoStop}%
\bibitem [{\citenamefont {Pariser}(1956)}]{pariser56}%
  \BibitemOpen
  \bibfield  {author} {\bibinfo {author} {\bibfnamefont {R.}~\bibnamefont
  {Pariser}},\ }\href@noop {} {\bibfield  {journal} {\bibinfo  {journal} {J.
  Chem. Phys.}\ }\textbf {\bibinfo {volume} {24}},\ \bibinfo {pages} {250}
  (\bibinfo {year} {1956})}\BibitemShut {NoStop}%
\bibitem [{\citenamefont {Cizek}\ \emph {et~al.}(1974)\citenamefont {Cizek},
  \citenamefont {Paldus},\ and\ \citenamefont {Hubac}}]{cizek74}%
  \BibitemOpen
  \bibfield  {author} {\bibinfo {author} {\bibfnamefont {J.}~\bibnamefont
  {Cizek}}, \bibinfo {author} {\bibfnamefont {J.}~\bibnamefont {Paldus}}, \
  and\ \bibinfo {author} {\bibfnamefont {I.}~\bibnamefont {Hubac}},\
  }\href@noop {} {\bibfield  {journal} {\bibinfo  {journal} {Int. J. Quantum
  Chem.}\ }\textbf {\bibinfo {volume} {8}},\ \bibinfo {pages} {951} (\bibinfo
  {year} {1974})}\BibitemShut {NoStop}%
\bibitem [{\citenamefont {Ramasesha}\ \emph {et~al.}(1996)\citenamefont
  {Ramasesha}, \citenamefont {Pati}, \citenamefont {Krishnamurthy},
  \citenamefont {Shuai},\ and\ \citenamefont {{Br\'{e}das}}}]{pati}%
  \BibitemOpen
  \bibfield  {author} {\bibinfo {author} {\bibfnamefont {S.}~\bibnamefont
  {Ramasesha}}, \bibinfo {author} {\bibfnamefont {S.~K.}\ \bibnamefont {Pati}},
  \bibinfo {author} {\bibfnamefont {H.~R.}\ \bibnamefont {Krishnamurthy}},
  \bibinfo {author} {\bibfnamefont {Z.}~\bibnamefont {Shuai}}, \ and\ \bibinfo
  {author} {\bibfnamefont {J.~L.}\ \bibnamefont {{Br\'{e}das}}},\ }\href@noop
  {} {\bibfield  {journal} {\bibinfo  {journal} {Phys. Rev. B}\ }\textbf
  {\bibinfo {volume} {54}},\ \bibinfo {pages} {7598} (\bibinfo {year}
  {1996})}\BibitemShut {NoStop}%
\bibitem [{\citenamefont {Prodhan}\ \emph {et~al.}()\citenamefont {Prodhan},
  \citenamefont {Mazumdar},\ and\ \citenamefont {Ramasesha}}]{suryo}%
  \BibitemOpen
  \bibfield  {author} {\bibinfo {author} {\bibfnamefont {S.}~\bibnamefont
  {Prodhan}}, \bibinfo {author} {\bibfnamefont {S.}~\bibnamefont {Mazumdar}}, \
  and\ \bibinfo {author} {\bibfnamefont {S.}~\bibnamefont {Ramasesha}},\
  }\href@noop {} {\enquote {\bibinfo {title} {Efficient symmetrized dmrg
  algorithm for low-lying excited states of conjugated carbon systems:
  Application to coronene, ovalene and 1,12-benzoperylene},}\ }\Eprint
  {http://arxiv.org/abs/1601.00053} {arXiv:1601.00053 [chem-ph]} \BibitemShut
  {NoStop}%
\bibitem [{\citenamefont {Hudson}\ and\ \citenamefont
  {Kohler}(1972)}]{Hudson-Kohler}%
  \BibitemOpen
  \bibfield  {author} {\bibinfo {author} {\bibfnamefont {B.~S.}\ \bibnamefont
  {Hudson}}\ and\ \bibinfo {author} {\bibfnamefont {B.~E.}\ \bibnamefont
  {Kohler}},\ }\href@noop {} {\bibfield  {journal} {\bibinfo  {journal} {Chem.
  Phys. Lett.}\ }\textbf {\bibinfo {volume} {14}},\ \bibinfo {pages} {299}
  (\bibinfo {year} {1972})}\BibitemShut {NoStop}%
\bibitem [{\citenamefont {Schulten}\ and\ \citenamefont
  {Karplus}(1972)}]{Schulten-Karplus}%
  \BibitemOpen
  \bibfield  {author} {\bibinfo {author} {\bibfnamefont {K.}~\bibnamefont
  {Schulten}}\ and\ \bibinfo {author} {\bibfnamefont {M.}~\bibnamefont
  {Karplus}},\ }\href@noop {} {\bibfield  {journal} {\bibinfo  {journal} {Chem.
  Phys. Lett.}\ }\textbf {\bibinfo {volume} {14}},\ \bibinfo {pages} {305}
  (\bibinfo {year} {1972})}\BibitemShut {NoStop}%
\bibitem [{\citenamefont {Chapra}\ and\ \citenamefont {Canale}(2012)}]{chapra}%
  \BibitemOpen
  \bibfield  {author} {\bibinfo {author} {\bibfnamefont {S.~C.}\ \bibnamefont
  {Chapra}}\ and\ \bibinfo {author} {\bibfnamefont {R.~P.}\ \bibnamefont
  {Canale}},\ }\href@noop {} {\emph {\bibinfo {title} {Numerical Methods for
  Engineers}}},\ \bibinfo {edition} {6th}\ ed.\ (\bibinfo  {publisher} {McGraw
  Hill Education (India) Private Limited},\ \bibinfo {address} {New Delhi},\
  \bibinfo {year} {2012})\ \bibinfo {note} {chapter 17}\BibitemShut {NoStop}%
\bibitem [{\citenamefont {Krotov}\ \emph {et~al.}(1997)\citenamefont {Krotov},
  \citenamefont {Lee},\ and\ \citenamefont {Louie}}]{louie97}%
  \BibitemOpen
  \bibfield  {author} {\bibinfo {author} {\bibfnamefont {Y.~A.}\ \bibnamefont
  {Krotov}}, \bibinfo {author} {\bibfnamefont {D.-H.}\ \bibnamefont {Lee}}, \
  and\ \bibinfo {author} {\bibfnamefont {S.~G.}\ \bibnamefont {Louie}},\
  }\href@noop {} {\bibfield  {journal} {\bibinfo  {journal} {Phys. Rev. Lett.}\
  }\textbf {\bibinfo {volume} {78}},\ \bibinfo {pages} {4245} (\bibinfo {year}
  {1997})}\BibitemShut {NoStop}%
\bibitem [{\citenamefont {Balents}\ and\ \citenamefont
  {Fisher}(1997)}]{balents97}%
  \BibitemOpen
  \bibfield  {author} {\bibinfo {author} {\bibfnamefont {L.}~\bibnamefont
  {Balents}}\ and\ \bibinfo {author} {\bibfnamefont {M.~P.~A.}\ \bibnamefont
  {Fisher}},\ }\href@noop {} {\bibfield  {journal} {\bibinfo  {journal} {Phys.
  Rev. B}\ }\textbf {\bibinfo {volume} {55}},\ \bibinfo {pages} {R11973}
  (\bibinfo {year} {1997})}\BibitemShut {NoStop}%
\bibitem [{\citenamefont {Deshpande}\ \emph {et~al.}(2009)\citenamefont
  {Deshpande}, \citenamefont {Chandra}, \citenamefont {Caldwell}, \citenamefont
  {Novikov}, \citenamefont {Hone},\ and\ \citenamefont
  {Bockrath}}]{deshpande09}%
  \BibitemOpen
  \bibfield  {author} {\bibinfo {author} {\bibfnamefont {V.~V.}\ \bibnamefont
  {Deshpande}}, \bibinfo {author} {\bibfnamefont {B.}~\bibnamefont {Chandra}},
  \bibinfo {author} {\bibfnamefont {R.}~\bibnamefont {Caldwell}}, \bibinfo
  {author} {\bibfnamefont {D.~S.}\ \bibnamefont {Novikov}}, \bibinfo {author}
  {\bibfnamefont {J.}~\bibnamefont {Hone}}, \ and\ \bibinfo {author}
  {\bibfnamefont {M.}~\bibnamefont {Bockrath}},\ }\href@noop {} {\bibfield
  {journal} {\bibinfo  {journal} {Science}\ }\textbf {\bibinfo {volume}
  {323}},\ \bibinfo {pages} {106} (\bibinfo {year} {2009})}\BibitemShut
  {NoStop}%
\end{thebibliography}%

\end{document}